\documentclass[reqno,12pt,a4paper]{amsart}

\usepackage{mathrsfs}
\usepackage{amssymb}
\usepackage{amsfonts}
\usepackage{latexsym}
\usepackage{amsthm}

\usepackage{graphicx}
\def\lb{\label}

\newcommand{\er}[1]{\textrm{(\ref{#1})}}

\begin{document}


\renewcommand{\theequation}{\arabic{section}.\arabic{equation}}
\theoremstyle{plain}
\newtheorem{theorem}{\bf Theorem}[section]
\newtheorem{lemma}[theorem]{\bf Lemma}
\newtheorem{corollary}[theorem]{\bf Corollary}
\newtheorem{proposition}[theorem]{\bf Proposition}
\newtheorem{definition}[theorem]{\bf Definition}
\newtheorem{condition}[theorem]{\bf Condition}
\newtheorem{remark}[theorem]{\bf Remark}

\def\a{\alpha}  \def\cA{{\mathcal A}}     \def\bA{{\bf A}}  \def\mA{{\mathscr A}}
\def\b{\beta}   \def\cB{{\mathcal B}}     \def\bB{{\bf B}}  \def\mB{{\mathscr B}}
\def\g{\gamma}  \def\cC{{\mathcal C}}     \def\bC{{\bf C}}  \def\mC{{\mathscr C}}
\def\G{\Gamma}  \def\cD{{\mathcal D}}     \def\bD{{\bf D}}  \def\mD{{\mathscr D}}
\def\d{\delta}  \def\cE{{\mathcal E}}     \def\bE{{\bf E}}  \def\mE{{\mathscr E}}
\def\D{\Delta}  \def\cF{{\mathcal F}}     \def\bF{{\bf F}}  \def\mF{{\mathscr F}}
\def\c{\chi}    \def\cG{{\mathcal G}}     \def\bG{{\bf G}}  \def\mG{{\mathscr G}}
\def\z{\zeta}   \def\cH{{\mathcal H}}     \def\bH{{\bf H}}  \def\mH{{\mathscr H}}
\def\e{\eta}    \def\cI{{\mathcal I}}     \def\bI{{\bf I}}  \def\mI{{\mathscr I}}
\def\p{\psi}    \def\cJ{{\mathcal J}}     \def\bJ{{\bf J}}  \def\mJ{{\mathscr J}}
\def\vT{\Theta} \def\cK{{\mathcal K}}     \def\bK{{\bf K}}  \def\mK{{\mathscr K}}
\def\k{\kappa}  \def\cL{{\mathcal L}}     \def\bL{{\bf L}}  \def\mL{{\mathscr L}}
\def\l{\lambda} \def\cM{{\mathcal M}}     \def\bM{{\bf M}}  \def\mM{{\mathscr M}}
\def\L{\Lambda} \def\cN{{\mathcal N}}     \def\bN{{\bf N}}  \def\mN{{\mathscr N}}
\def\m{\mu}     \def\cO{{\mathcal O}}     \def\bO{{\bf O}}  \def\mO{{\mathscr O}}
\def\n{\nu}     \def\cP{{\mathcal P}}     \def\bP{{\bf P}}  \def\mP{{\mathscr P}}
\def\r{\rho}    \def\cQ{{\mathcal Q}}     \def\bQ{{\bf Q}}  \def\mQ{{\mathscr Q}}
\def\s{\sigma}  \def\cR{{\mathcal R}}     \def\bR{{\bf R}}  \def\mR{{\mathscr R}}
\def\S{\Sigma}  \def\cS{{\mathcal S}}     \def\bS{{\bf S}}  \def\mS{{\mathscr S}}
\def\t{\tau}    \def\cT{{\mathcal T}}     \def\bT{{\bf T}}  \def\mT{{\mathscr T}}
\def\f{\phi}    \def\cU{{\mathcal U}}     \def\bU{{\bf U}}  \def\mU{{\mathscr U}}
\def\F{\Phi}    \def\cV{{\mathcal V}}     \def\bV{{\bf V}}  \def\mV{{\mathscr V}}
\def\P{\Psi}    \def\cW{{\mathcal W}}     \def\bW{{\bf W}}  \def\mW{{\mathscr W}}
\def\o{\omega}  \def\cX{{\mathcal X}}     \def\bX{{\bf X}}  \def\mX{{\mathscr X}}
\def\x{\xi}     \def\cY{{\mathcal Y}}     \def\bY{{\bf Y}}  \def\mY{{\mathscr Y}}
\def\X{\Xi}     \def\cZ{{\mathcal Z}}     \def\bZ{{\bf Z}}  \def\mZ{{\mathscr Z}}
\def\O{\Omega}
\def\th{\theta}

\newcommand{\gA}{\mathfrak{A}}
\newcommand{\gB}{\mathfrak{B}}
\newcommand{\gC}{\mathfrak{C}}
\newcommand{\gD}{\mathfrak{D}}
\newcommand{\gE}{\mathfrak{E}}
\newcommand{\gF}{\mathfrak{F}}
\newcommand{\gG}{\mathfrak{G}}
\newcommand{\gH}{\mathfrak{H}}
\newcommand{\gI}{\mathfrak{I}}
\newcommand{\gJ}{\mathfrak{J}}
\newcommand{\gK}{\mathfrak{K}}
\newcommand{\gL}{\mathfrak{L}}
\newcommand{\gM}{\mathfrak{M}}
\newcommand{\gN}{\mathfrak{N}}
\newcommand{\gO}{\mathfrak{O}}
\newcommand{\gP}{\mathfrak{P}}
\newcommand{\gQ}{\mathfrak{Q}}
\newcommand{\gR}{\mathfrak{R}}
\newcommand{\gS}{\mathfrak{S}}
\newcommand{\gT}{\mathfrak{T}}
\newcommand{\gU}{\mathfrak{U}}
\newcommand{\gV}{\mathfrak{V}}
\newcommand{\gW}{\mathfrak{W}}
\newcommand{\gX}{\mathfrak{X}}
\newcommand{\gY}{\mathfrak{Y}}
\newcommand{\gZ}{\mathfrak{Z}}

\newcommand{\gm}{\mathfrak{m}}
\newcommand{\gn}{\mathfrak{n}}
\newcommand{\gf}{\mathfrak{f}}
\newcommand{\gh}{\mathfrak{h}}
\newcommand{\mg}{\mathfrak{g}}

\def\ve{\varepsilon}   \def\vt{\vartheta}    \def\vp{\varphi}    \def\vk{\varkappa}

\def\Z{{\mathbb Z}}    \def\R{{\mathbb R}}   \def\C{{\mathbb C}}    \def\K{{\mathbb K}}
\def\T{{\mathbb T}}    \def\N{{\mathbb N}}   \def\dD{{\mathbb D}}


\def\la{\leftarrow}              \def\ra{\rightarrow}            \def\Ra{\Rightarrow}
\def\ua{\uparrow}                \def\da{\downarrow}
\def\lra{\leftrightarrow}        \def\Lra{\Leftrightarrow}


\def\lt{\biggl}                  \def\rt{\biggr}
\def\ol{\overline}               \def\wt{\widetilde}
\def\no{\noindent}


\let\ge\geqslant                 \let\le\leqslant
\def\lan{\langle}                \def\ran{\rangle}
\def\/{\over}                    \def\iy{\infty}
\def\sm{\setminus}               \def\es{\emptyset}
\def\ss{\subset}                 \def\ts{\times}
\def\pa{\partial}                \def\os{\oplus}
\def\om{\ominus}                 \def\ev{\equiv}
\def\iint{\int\!\!\!\int}        \def\iintt{\mathop{\int\!\!\int\!\!\dots\!\!\int}\limits}
\def\el2{\ell^{\,2}}             \def\1{1\!\!1}
\def\sh{\sharp}
\def\wh{\widehat}
\def\bs{\backslash}

\def\sh{\mathop{\mathrm{sh}}\nolimits}
\def\Area{\mathop{\mathrm{Area}}\nolimits}
\def\arg{\mathop{\mathrm{arg}}\nolimits}
\def\const{\mathop{\mathrm{const}}\nolimits}
\def\det{\mathop{\mathrm{det}}\nolimits}
\def\diag{\mathop{\mathrm{diag}}\nolimits}
\def\diam{\mathop{\mathrm{diam}}\nolimits}
\def\dim{\mathop{\mathrm{dim}}\nolimits}
\def\dist{\mathop{\mathrm{dist}}\nolimits}
\def\Im{\mathop{\mathrm{Im}}\nolimits}
\def\Iso{\mathop{\mathrm{Iso}}\nolimits}
\def\Ker{\mathop{\mathrm{Ker}}\nolimits}
\def\Lip{\mathop{\mathrm{Lip}}\nolimits}
\def\rank{\mathop{\mathrm{rank}}\limits}
\def\Ran{\mathop{\mathrm{Ran}}\nolimits}
\def\Re{\mathop{\mathrm{Re}}\nolimits}
\def\Res{\mathop{\mathrm{Res}}\nolimits}
\def\res{\mathop{\mathrm{res}}\limits}
\def\sign{\mathop{\mathrm{sign}}\nolimits}
\def\span{\mathop{\mathrm{span}}\nolimits}
\def\supp{\mathop{\mathrm{supp}}\nolimits}
\def\Tr{\mathop{\mathrm{Tr}}\nolimits}
\def\BBox{\hspace{1mm}\vrule height6pt width5.5pt depth0pt \hspace{6pt}}
\def\as{\text{as}}
\def\all{\text{all}}
\def\where{\text{where}}
\def\Dom{\mathop{\mathrm{Dom}}\nolimits}


\newcommand\nh[2]{\widehat{#1}\vphantom{#1}^{(#2)}}
\def\dia{\diamond}

\def\Oplus{\bigoplus\nolimits}



\def\qqq{\qquad}
\def\qq{\quad}
\let\ge\geqslant
\let\le\leqslant
\let\geq\geqslant
\let\leq\leqslant
\newcommand{\ca}{\begin{cases}}
\newcommand{\ac}{\end{cases}}
\newcommand{\ma}{\begin{pmatrix}}
\newcommand{\am}{\end{pmatrix}}
\renewcommand{\[}{\begin{equation}}
\renewcommand{\]}{\end{equation}}
\def\eq{\begin{equation}}
\def\qe{\end{equation}}
\def\[{\begin{equation}}
\def\bu{\bullet}

\newcommand{\fr}{\frac}
\newcommand{\tf}{\tfrac}

\title[Mc'Kean's transformation for 3-rd order operators]
{Mc'Kean's transformation for 3-rd order operators}

\date{\today}
\author[Andrey Badanin]{Andrey Badanin}
\author[Evgeny Korotyaev]{Evgeny L. Korotyaev}
\address{Saint-Petersburg
State University, Universitetskaya nab. 7/9, St. Petersburg,
199034 Russia,
an.badanin@gmail.com,\  a.badanin@spbu.ru,\
korotyaev@gmail.com,\  e.korotyaev@spbu.ru}


\maketitle

\begin{abstract}
We consider a non-self-adjoint third order operator $(y''+py)'+py'+qy$
with 1-periodic coefficients $p,q$. This operator is the L-operator
in the Lax pair for the good Boussinesq equation on the circle.
In 1981, McKean introduced a transformation that reduces
the spectral problem for this operator to a spectral problem
for the Hill operator with a potential that depends analytically on the energy.
In the present paper we are studying this transformation.
\end{abstract}

\section{Introduction and main results}
\setcounter{equation}{0}

\subsection{Introduction}
We consider a non-self-adjoint operator $\cL=\cL(\p)$ acting on $L^2(\R)$ and
given by
\[
\lb{Hpq}
\cL y=(y''+py)'+py'+qy,
\]
where the 1-periodic coefficients $p,q$ satisfy
$$
\p=(p,q)\in \cH\os\cH,
$$
the Hilbert space $\cH=L_\R^2(\T),\T=\R/\Z$, equipped
with the norm $\|f\|^2=\int_0^1|f(x)|^2dx$.
Let $\cH_\C$ be the complexification of the real space $\cH$.
The operator is defined on the domain
\[
\lb{cDH}
\begin{aligned}
\Dom(\cL )=\Big\{y\in L^2(\R):(y''+py)'+py'+qy\in L^2(\R),
y'',(y''+py)'\in L^1(\R)\Big\}.
\end{aligned}
\]

The operator $\cL $ is the $L$-operator in the Lax pair
for the good Boussinesq equation on the circle.
The inverse problem for this operator was studied by McKean's \cite{McK81}.
The solution of the inverse problem uses a transformation that reduces the operator $\cL $
to the Schrodinger operator with a periodic energy-dependent potential.
Studying this transformation is the main goal of the present work.

Recall that in order to recover the potential of the Schr\"odinger operator
with a periodic potential, three sequences are required:

1) Sequence of eigenvalues of a 2-periodic problem.

2) Sequence of eigenvalues of the Dirichlet problem.

3) Sequence of +/- signs.

We showed in \cite{BK24} that in order to recover the coefficients of the operator $\cL $,
three sequences are also required:

1) Sequence of branch points of the Riemann surface of the operator.

2) Sequence of eigenvalues of the 3-point Dirichlet problem.

3) Sequence of +/- signs.

\no We prove in the present paper that in the case of small coefficients
McKean's transformation establishes
a correspondence between these three sequences for the operator $\cL $
and the Schr\"odinger operator with an energy-dependent potential.

Note that since our operator is non-self-adjoint, the branch points
of the Riemann surface and the spectrum of the 3-point problem
can be non-real. The spectra of the 2-periodic problem and
the Dirichlet problem for the Schrodinger operator
with an energy-dependent potential can also be complex.
In our article \cite{BK21} we proved that in the case
of a sufficiently small potential the spectrum of the Schr\"odinger
operator is real. Using this fact, in the present paper we prove
that for small coefficients the branch points of the Riemann surface
and the spectrum of the 3-point problem of the operator $\cL $ are real.

The inverse spectral problem for the Schr\"odinger operator with the periodic
potential in the class of potentials from $L^2(\T)$
is proved by Marchenko and Ostrovskii \cite{MO75} and
Korotyaev \cite{K99}.
Some inverse results for the 3-order operator $\cL $ are obtained by
McKean \cite{McK81}. There, for the case of small smooth coefficients,
a uniqueness theorem
for recovering the potential from spectral data was proved.
The problem in the certain classes of non-smooth coefficients
is more difficult. Some results in this direction are presented in our article
\cite{BK24}.

\subsection{Second order operators}
Recall the results for the Schr\"odinger operator $-f''+Vf$ with the potential
$V\in\cH$, acting on $L^2(\R)$. The spectrum is pure absolutely
continuous and consists of the intervals $[E_{n-1}^+,E_n^-]$
separated by gaps, where $E_0^+<E_1^-\le E_1^+<...$ are the
eigenvalues of the 2-periodic problem $f(0)=f(2),f'(0)=f'(2)$.
The spectrum of the Dirichlet problem $f(0)=f(1)=0$
consists of the simple eigenvalues  $\gm_1<\gm_2<...$.
This eigenvalues are zeros of the entire function $\vp(1,\cdot)$,
where $\vp(x,\l)$ is the fundamental solution of the equation
\[
\lb{hilleq}
-f''+Vf=\l f,
\]
satisfying the initial conditions
$\vp|_{x=0}=0,\vp'|_{x=0}=1$.
The eigenvalues satisfy $\gm_n\in[E_n^-,E_n^+]$ for all $n\in\N$.
Introduce the {\it norming constants}
$\gh_{sn}:\cH\to\R$ by
\[
\lb{ncschr}
\gh_{sn}(V)=2\pi n\log|\vp'(1,\gm_n(V))|,\qq\log 1=0,\qq n\in\N.
\]
Introduce the space $S$ of all real, strictly increasing
sequences $(s_1,s_2,...)$ such that $(s_n-(\pi n)^2)\in\ell^2$.
The following result holds true, see, e.g., \cite{PT87}.
\medskip

{\it
The mapping $V\to (\gm_n(V),\gh_{sn}(V))$
is a real analytic isomorphism between $\cH$ and $S\os\ell^2$.
}

\medskip

\no {\bf Remark.} We solve the corresponding inverse problem for 3-rd order
operator under the 3-point Dirichlet conditions in \cite{BK24x}.

\medskip

The inverse problem of recovering the potential
$V$ from the periodic spectrum was solved by Korotyaev \cite{K99}.
Introduce the mapping $\mg:\cH\to \ell^2\os\ell^2$ from \cite{K99} by
\[
\lb{defwtg}
\mg(V)=(\mg_n(V))_{n\in\N},\qq V\in\cH,
\]
where
$$
\mg_n=(\mg_{cn},\mg_{sn})\in\R^2,
\qq |\mg_n|={E_n^+-E_n^-\/2},
$$
the vectors $(\mg_{cn},\mg_{sn})$ are given by
\[
\lb{defwtgcn}
\mg_{cn}={E_n^++E_n^-\/2}-\gm_n,\qq
\]
\[
\lb{defwtgsn}
\mg_{sn}=\Big|{(E_n^+-E_n^-)^2\/4}-\mg_{cn}^2\Big|^{1\/2}\sign \gh_{sn}.
\]
Introduce the Fourier transformations
$\F_{cn},\F_{sn}:\cH\to\R,n\in\N$, by
$$
f_{cn}=\F_{cn}f=\int_0^1f(x)\cos 2\pi nxdx,\ \
f_{sn}=\F_{sn}f=\int_0^1f(x)\sin 2\pi nxdx.
$$
and
$
\F_n=(\F_{cn},\F_{sn})_{n\in\N}.
$
The following results were proved by Korotyaev \cite{K99}.

\medskip

{\it The mapping $\mg$, given by
\er{defwtg}--\er{defwtgsn},
is a real analytic isomorphism between $\cH$ and $\ell^2\os\ell^2$.
Moreover, if $n\to+\iy$, then the functions $\mg_n$ satisfy
$$
\mg_n(V)=(V_{cn},V_{sn})+O(n^{-1}),
\qqq
{\pa\mg_n\/\pa V}=\F_n+O(n^{-1}).
$$
}

\medskip

\no {\bf Remark.} We solve the corresponding inverse problem for 3-rd order
operator in \cite{BK24}.

\subsection{McKean's transformation}
Consider the differential equation
\[
\lb{1b}
(y''+py)'+py'+q y=\l y,\qqq\l\in\C,
\]
in the class
$$
C^{[2]}(\R)=\{y:y,y',y^{[2]}\in AC_{loc}(\R)\},
$$
where
\[
\lb{defmL}
y^{[2]}=y''+ py.
\]
McKean's transformation is based on the following simple result.

\begin{lemma}
\lb{Lmtr2od1}
Let $p,q\in L_{loc}^1(\R),\l\in\C$.
Assume that Eq.~\er{1b}
has a solution $\eta\in C^{[2]}(\R)$
such that $\eta(x)\ne 0$ for all $x\in\R$.
Let, in addition, $y\in C^{[2]}(\R)$ be a solution to Eq.~\er{1b}.
Let the function $f$ has the form
\[
\lb{deff1}
f=\eta^{3\/2}\Big({y\/\eta}\Big)'={y'\eta-y\eta'\/\sqrt{\eta}}.
\]
Then $f,f',f''+{p\/2}f\in AC_{loc}(\R)$, and $f$ satisfies the equation
\[
\lb{2oequd1}
-f''+Vf=E f,
\]
where
\[
\lb{defEen}
E={3z^2\/4}={3 \l^{2\/3}\/4},\qq
\l=\Big({4\/3}E\Big)^{3\/2},\qq z=\l^{1\/3},\qq\arg z\in(-{\pi\/3},{\pi\/3}],
\qq\arg\l\in(-\pi,\pi],
\]
the energy-dependent potential
$V$ has the form
\[
\lb{da1+}
V=E-{p\/2}-{3\/2}{\eta^{[2]}\/\eta}+{3\/4}\Big({\eta'\/\eta}\Big)^2.
\]
Moreover, $V+{p\/2}\in AC_{loc}(\R)$.
\end{lemma}

\no {\bf Remark.}
1) In the unperturbed  case $\p=0$ we have $V=0$.

2) If the coefficients $p,q$ are 1-periodic and $\p\in\cH\os\cH$, then
the potential $V$ is also 1-periodic with respect to the variable $x$,
and $V+{p\/2}\in AC(\T)$.

3) Eq.~\er{2oequd1} was the subject of our article \cite{BK21}.

\subsection{Ramifications}
Now we consider the periodic coefficients.
Let $\p\in\cH\os\cH$.
Introduce the {\it fundamental solutions} $\vp_1, \vp_2, \vp_3$ of Eq.~\er{1b}
satisfying the conditions $\vp_j^{[k-1]}|_{x=0}=\d_{jk}$, $j,k=1,2,3$,
where $y^{[0]}=y$, $y^{[1]}=y'$, $y^{[2]}=y''+py$.
Each of the functions $\vp_j(x,\cdot),j=1,2,3,x\in[0,1]$ is entire
and real on $\R$.
We define
the {\it monodromy matrix} by
\[
\lb{defmm}
M(\l)=\big(\vp_j^{[k-1]}(1,\l)\big)_{j,k=1}^3.
\]
The matrix-valued function $M$ is entire.
The characteristic polynomial $D$ of the monodromy matrix is given by
\[
\lb{1c} D(\t,\l)
=\det(M(\l)-\t \1_{3}),\qq (\t,\l)\in\C^2,
\]
here and below $\1_N$ is the $N\ts N$ identity matrix.
An eigenvalue  of $M$ is called a {\it multiplier}, it is a
zero of the polynomial $D(\cdot,\l)$.
Each $M$ has exactly $3$ (counting with
multiplicities) multipliers $\t_j,j=1,2,3$, which  satisfy
$
\t_1\t_2\t_3=1.
$
In particular, each $\t_j\ne 0$ for all $\l\in\C$.

Introduce the Floquet solutions $\gf_k,k=1,2,3$,
of Eq.~\er{1b} satisfying the conditions
\[
\lb{Fs1}
\gf_k(x+1,\l)=\t_k(\l)\gf_k(x,\l),\qqq \gf_k(0,\l)=1.
\]

Three roots $\t_1, \t_2$ and $\t_3$ of the equation $D(\cdot,\l)=0$
constitute three distinct branches of some analytic function $\t$
of the variable $\l$
that has only algebraic singularities in $\C$.
Thus the function $\t$ is
analytic on a 3-sheeted Riemann surface $\cR$, Fig.~\ref{FigrsZEL}~c),
see \cite{BK24xxx}.
Moreover, for each $x\in\R$
the functions $\gf_{1}(x,\cdot),\gf_{2}(x,\cdot),\gf_{3}(x,\cdot)$
constitute three branches of a function
meromorphic on the surface $\cR$.

There are only a finite number of the algebraic singularities
(ramifications)  in
any bounded domain. In order to describe these points we
introduce
the {\it discriminant} $\r$
of the polynomial
$D(\cdot,\l)$ by
\[
\lb{r}
\r=(\t_1-\t_2)^2(\t_1-\t_3)^2(\t_2-\t_3)^2.
\]
The function $\r(\l)$ is entire.
A zero of $\r$ is a {\it ramification}
of the Riemann surface $\cR$.

If $\p=0$, then the function $\r$ and its zeros $r_{n}^{0,\pm}$
have the form
\[
\lb{ro0}
\r_0=64\sin^2{\sqrt 3 z\/2}\sin^2{\sqrt 3 \o z\/2}
\sin^2{\sqrt 3 \o^2z\/2},\qq
r_{n}^{0,+}=r_{n}^{0,-}=\Big({2\pi n\/\sqrt3}\Big)^3,\ \ n\in\Z,
\qq \o=e^{i{2\pi\/3}}.
\]

If the coefficients $p,q$ are 1-periodic and $\p\in\cH\os\cH$, then
the potential $V$, given by \er{da1+}, is also 1-periodic with respect to the variable $x$,
and $V+{p\/2}\in AC(\T)$.
Consider the 2-periodic problem
\[
\lb{fbc}
f(2)=f(0),\qqq f'(2)=f'(0),
\]
for Eq.~\er{2oequd1}. Let each of the potentials $V(x,\cdot),x\in\R$, be analytic on
some neighborhood of $E_o\in\C$ and let
the problem \er{2oequd1}, \er{fbc} with $E=E_o$ has a non-trivial solution.
Then $E_o$ is an eigenvalue of this problem, see \cite{BK21}.

We prove the following result about the ramification.

\begin{theorem}
\lb{Lmtpevsch}
Let $\p\in\cH_\C^2$. Assume that there exists
a simple multiplier $\t$ for Eq.~\er{1b}
with some $\l_o\in\C$ and the
corresponding Floquet solution $\eta\in C^{[2]}(\R)$ does not vanish for all $x\in\R$.
Let McKean's transformation be given by \er{deff1}--\er{da1+}.
Then each of the potentials $V(x,\cdot),x\in\R$, is analytic on some neighborhood
of $E_o={3\/4}\l_o^{2\/3}$.
Moreover, $\l_o$ is a ramification of Eq.~\er{1b} iff  $E_o$
is an eigenvalue of the 2-periodic problem \er{2oequd1}, \er{fbc}.
\end{theorem}

\no {\bf Remark.}
The idea of proving this result was proposed by McKean \cite{McK81}.
Following \cite{McK81}, in the proof we use two substitutions in \er{deff1}:
we take $g=\gf_1$ in the first substitution, and we take $g=\gf_2$ in the second one,
here $\gf_1,\gf_2$ are the Floquet solutions corresponding the multipliers $\t_1,\t_2$,
respectively.

\subsection{Three-point Dirichlet problem}
We also consider the operator $\cL_{dir}$ acting on $L^2(0,2)$ and
given by
\[
\lb{Hdpq}
\cL _{dir}y=(y''+py)'+py'+qy,\qqq y(0)=y(1)=y(2)=0.
\]
The spectrum $\s(\cL _{dir})$ of $\cL _{dir}$ is pure discrete
and satisfies
\[
\lb{spec}
\s(\cL _{dir})=\{\l\in\C:\mB(\l)=0\},
\]
where $\mB$ is the entire function given by
\[
\lb{defsi}
\mB(\l)=\det\ma\vp_2(1,\l)&\vp_3(1,\l)\\
\vp_2(2,\l)&\vp_3(2,\l)\am.
\]

In the unperturbed case $\p=0$ the function $\mB$ has the form
\[
\lb{mB0}
\mB_0(\l)=-{8\/3\sqrt3 \l}\sin{\sqrt 3 z\/2}
\sin{\sqrt 3 \o z\/2} \sin{\sqrt 3 \o^2z\/2}.
\]
All eigenvalues are simple, real, and
have the form
\[
\lb{unpev}
\m_{n}^o=\Big({2\pi n\/\sqrt3}\Big)^3,\qq n\in\Z\sm\{0\}.
\]

Consider the Dirichlet problem
\[
\lb{dbc}
f(0)=f(1)=0,
\]
for Eq.~\er{2oequd1}.
Let each of the potentials $V(x,\cdot),x\in\R$, be analytic on
some neighborhood of $E_o\in\C$ and let
the problem \er{2oequd1}, \er{dbc} with $E=E_o$ has a non-trivial solution.
Then $E_o$ is an eigenvalue of this problem, see \cite{BK21}.

Consider the transpose operator
\[
\lb{trOp}
\wt\cL y=-(y''+py)'-py'+q y.
\]
Note that the operators $\cL(\p)=-\wt\cL(\p_*)$ and $-\cL(\p_*^-)$
are unitarily equivalent, where
\[
\lb{epsstar}
\p_*=(p,-q),
\qq
\p^-(x)=\p(1-x)
,\qq \p_*^-(x)=\p_*(1-x)
,\qq x\in\R.
\]
In particular, if $y(\cdot,\l,\p),(\l,\p)\in\C\ts\cH\os\cH$, is a solution of
Eq.~\er{1b} for some $\l\in\C$, then
$\tilde y(\cdot,\l,\p)=y(\cdot,-\l,\p_*)$ is
a solution of the equation
\[
\lb{1btr}
-(\tilde y''+p\tilde y)'-p\tilde y'+q \tilde y=\l \tilde y.
\]
The function $\tilde y=y'\gf_3-y\gf_3'$ in the numerator of
the substitution \er{deff1} satisfies Eq.~\er{1btr},
see Lemma~\ref{Lmsolconjeq}.

We also consider the operator $\wt\cL_{dir}$ which is a restriction
of the operator $\wt\cL$ onto the functions satisfying
the 3-point conditions $y(0)=y(1)=y(2)=0$.

We prove the following result about the 3-point eigenvalue.

\begin{theorem}
\lb{LmtpevschDir}
Let $\p\in\cH_\C^2$. Assume that there exists
a simple multiplier $\t$ for Eq.~\er{1b}
with some $\l\in\C$ and the
corresponding Floquet solution $\eta\in C^{[2]}(\R)$ does not vanish for all $x\in\R$.
Let McKean's transformation be given by the identities
\er{deff1}--\er{da1+}.

Let $\l$ be an eigenvalue of the 3-point Dirichlet problem
for Eq.~\er{1btr} and let $\tilde y$ be the corresponding eigenfunction.
If, in addition, $\eta^{[2]}(0)\ne 0$,
then
$E={3\/4}\l^{2\/3}$ is an eigenvalue of the Dirichlet problem
\er{2oequd1}, \er{dbc} and $f={\tilde y\/\sqrt \eta}$
is the corresponding eigenfunction.

Conversely, let $E$ be an eigenvalue of the Dirichlet problem
\er{2oequd1}, \er{dbc} and let $f$ be the corresponding eigenfunction.
Then $\l=({4\/3}E)^{3\/2}$ is an eigenvalue of the
3-point Dirichlet problem for Eq.~\er{1btr}
 and $\tilde y=f\sqrt \eta$ is
the corresponding eigenfunction.
\end{theorem}

\no {\bf Remark.} In the proof we use the substitution \er{deff1},
where $y'\eta-y\eta'$ is the eigenfunction $\wt y$ of the 3-point problem for
Eq.~\er{1btr}.

\subsection{Localization of the ramifications}
Theorems~\ref{Lmtpevsch} and \ref{LmtpevschDir} prove
that McKean's transformation introduced in Lemma~\ref{Lmtr2od1}
transforms the ramification to the eigenvalue of the 2-periodic problem,
and the 3-point eigenvalue to the eigenvalue of the Dirichlet problem
for the Schr\"odinger operator.
Next we consider the case of small coefficients.
In this case, we prove a global one-to-one correspondence
between the set of positive (excluding $r_0^\pm$) ramifications
and the set of 2-periodic eigenvalues of the Schr\"odinger operator,
and a similar correspondence between the set of positive 3-point eigenvalues
and the set of  the Dirichlet eigenvalues for the Schr\"odinger
operator. The same correspondences are established
for negative ramifications and 3-point eigenvalues.
Moreover, in this case it is possible to prove the reality of
ramifications (except $r_0^\pm$) and 3-point eigenvalues.

Introduce the domains $\cD_n, n\in\Z$, by
\[
\lb{DomcD}
\cD_{0}=\Big\{\l\in\C:|\l|<1\Big\},\ \
\cD_{n}=\Big\{\l\in\C:\big|z-{2\pi n\/\sqrt3}\big|
<1\Big\},\ \
\cD_{-n}=\big\{\l\in\C:-\l\in\cD_n\big\},\ \ n\in\N.
\]
Always below $\ve_0>0$ is a fixed small enough number.
Introduce the ball
$$
\cB(\ve_0)=\{\p\in\cH\os\cH:\|\p\|<\ve\},
$$
and the corresponding ball $\cB_\C(\ve_0)$ in $\cH_\C\os\cH_\C$.
In \cite{BK24xxx} and \cite{BK24xx} we prove the following results.
\medskip

{\it
Let $\p\in\cB(\ve_0)$.
Then

i) There are exactly two (counting with multiplicity)
ramifications $r_n^\pm$ in each domain $\cD_n,n\in\Z$
and there are no ramifications in $\C\sm\cup_{n\in\Z}\cD_n$.
Moreover,
\[
\lb{simram}
r_n^\pm(\p)=-r_{-n}^\mp(\p_*)=-r_{-n}^\mp(\p_*^-)=r_n^\pm(\p^-)\qq
\qq\forall\ \ n\in\Z.
\]

ii) There exists the multiplier $\t_3$, satisfying the estimate
\[
\lb{lom}
|\t_3(\l)-e^{z}|\le {C\|\p\|\/|z|}|e^{z}|,
\]
for all $\l\in\mD_3$ and for some $C>0$, where
\[
\lb{defmD3}
\mD_3=\C_+\sm\cup_{n\in\{0\}\cup\N}\ol{\cD_{-n}}.
\]
Moreover, if  $\l\in\mD_3$,
then $\t_3(\l)$ is the simple multiplier
and the corresponding Floquet solution $\gf_3(x,\l)$ does not vanish
for all $x\in\R$.
Each function $\t_3$ and  $\gf_3(x,\cdot),x\in\R$,
is analytic on $\mD_3$,
$\t_3$ is real for real $\l$, and satisfies
\[
\lb{t3>0}
\t_3(\l)>0,\qq \forall\ \ \l>1.
\]
}

Introduce images $\cS$ and $\cS_n$ of the domains
$\mD_3$ and $\cD_n$  in the $E$-plane by
\[
\lb{mDBe}
\cS=\Big\{E\in\C:\l=\Big({4\/3}E\Big)^{3\/2}\in\mD_3\Big\},
\]
and
\[
\lb{defOmegan}
\cS_n=\Big\{E\in\C:\l=\Big({4\/3}E\Big)^{3\/2}\in\cD_n\Big\}
=\Big\{E\in\C:|\sqrt E-\pi n|<{\sqrt 3\/2}\Big\},\qq
n\in\N.
\]
Due to the statement iii), if  $\p\in\cB(\ve_0)$, then
for any $\l\in\mD_3$ we may take $\eta=\gf_3$ in
McKean's transformation \er{deff1}--\er{da1+}.
Then the potential $V(x,E),x\in\R$, is analytic with respect
to $E$ on the domain $\cS$.

In order to consider the Schr\"odinger operator with the energy-dependent
potential we assume $p',q\in L^2(\T)$.
Introduce the Sobolev space $\cH_1=H^1(\T)$.
Introduce the norm $\|\p\|_1$ in the space $\cH_1\os\cH$ by
$$
\|\p\|_1^2=\|p\|^2+\|p'\|^2+\|q\|^2.
$$
Let $\cH_{1,\C}$ be the complexification of the real space $\cH_1$.
Introduce the ball
$$
\cB_1(\ve_0)=\{\p\in\cH_1\os\cH:\|\p\|<\ve\},
$$
and the corresponding ball $\cB_{1,\C}(\ve_0)$ in $\cH_{1,\C}\os\cH_\C$.
We prove the following results about the ramifications.

\begin{theorem}
\lb{ThMcKtr}
Let  $\p\in\cB_{1,\C}(\ve_0)$. Then

i) There are exactly two
(counted with multiplicity)  eigenvalues $E_n^\pm$
of the 2-periodic problem \er{2oequd1},\er{fbc} in
each domain $\cS_n,n\in\N$, and there are no eigenvalues
 in the domain $\cS\sm\cup_{n\in\N}\cS_{n}$.
If, in addition, $\p$ is real, then
the eigenvalues $E_n^\pm,n\in\N$,
are real and satisfy
\[
\lb{labEjpm}
2<E_1^-\le E_1^+<E_2^-\le E_2^+<...
\]

ii) The ramifications  $r_n^\pm$ satisfy
\[
\lb{relr3p2o}
E_n^\pm(\p)=E_n^\pm(\p^-)={3\/4}(r_n^\pm(\p))^{2\/3}\qq\forall\ \ n\in\N.
\]
\[
\lb{relr3p2o-}
E_n^\pm(\p_*)=E_n^\pm(\p_*^-)={3\/4}(-r_{-n}^\mp(\p))^{2\/3}\qq\forall\ \ n\in\N.
\]
If $\p$ is real, then all $r_n^\pm,n\in\Z\sm\{0\}$, are real
and satisfy
$$
...<r_{-1}^-\le r_{-1}^+<r_1^-\le r_1^+<r_2^-\le r_2^+<...
$$

\end{theorem}

\no {\bf Remark.}
Proof of Theorem~\ref{ThMcKtr} is based on our results from
\cite{BK21}.

\subsection{Localization of 3-point eigenvalues and norming constants}

In \cite{BK24xx} we prove the following results.
\medskip

{\it
 Let $\p\in\cB_\C(\ve_0)$.
Then there is exactly one simple
eigenvalue $\m_n$ of the operator $\cL_{dir}$
and  exactly one simple
eigenvalue $\wt\m_n$ of the operator $\wt\cL_{dir}$
in each domain $\cD_n,n\in\Z\sm\{0\}$.
Moreover,
\[
\lb{symev}
\m_n(\p)=-\m_{-n}(\p_*^-)=-\wt\m_{-n}(\p_*)=\wt\m_n(\p^-),\qq\forall\ \ n\in\Z\sm\{0\}.
\]
}

In order to define norming constants  we need to consider the transpose
operator $\wt\cL_{dir}$.
If $\p\in\cB_\C(\ve_0)$, then in each
domain $\cD_n,n\in\Z_0$, there is exactly one real simple eigenvalue $\wt\m_n$ of
this operator.
Let $\wt y_n(x)$ be the corresponding eigenfunction such that $\wt y_n'(0)=1$.
We introduce the norming constants $h_{sn}, n\in \N$, by
\[
\lb{defnf}
 h_{sn} =8(\pi n)^2\log |\wt y_n'(1)\t_3^{-{1\/2}}(\wt\m_n)|, \qq
\t_3^{1\/2}(\wt\m_n)>0.
\]

We formulate our theorem about the 3-point eigenvalues and  norming constants.

\begin{theorem}
\lb{ThMcKtrDir}
Let  $\p\in\cB_{1,\C}(\ve_0)$. Then

i) There is exactly one simple eigenvalue $\gm_n$
of the Dirichlet problem \er{2oequd1}, \er{dbc}
in each domain $\cS_n,n\in\N$, and there are no eigenvalues
 in the domain $\cS\sm\cup_{n\in\N}\cS_{n}$.
If, in addition, $\p$ is real, then all eigenvalues $\gm_n,n\in\N$,
are real and satisfy
\[
\lb{gmj}
\gm_1<\gm_2<...,\qq
\gm_n\in[E_n^-,E_n^+],\qq\forall\ \ n\in\N.
\]

ii) The eigenvalues $\m_n$ of the 3-point problem satisfy
\[
\lb{relr3p2oDir}
\gm_n(\p^-)={3\/4}(\m_n(\p))^{2\/3},\qq\forall\ \ n\in\N,
\]
\[
\lb{relevdir-}
\gm_n(\p)={3\/4}\big(\wt\m_n(\p)\big)^{2\/3}
={3\/4}\big(-\m_{-n}(\p_*)\big)^{2\/3},\qq\forall\ \ n\in\N.
\]
If $\p$ is real, then all $\m_n,n\in\Z\sm\{0\}$, are real and satisfy
\[
\lb{altmurn}
\mu_{n}\in[r_{n}^-,r_{n}^+].
\]

iii) The norming constants satisfy
\[
\lb{nc2o}
\gh_{sn}={h_{sn}\/4\pi n},\qq\forall\ \ n\in\N.
\]

\end{theorem}

\no {\bf Remark.} The relation \er{altmurn} was obtained in \cite{McK81}
for the case of small $p,q\in C^\iy(\T)$ in a different way,
without using McKean's transformation.
We use McKean's transformation in this paper, which
gives a simpler proof for a wider class of coefficients.

\section{McKean's transformation}
\setcounter{equation}{0}

\subsection{Transformation of Eq.~\er{1b}}
We prove our first results about McKean's transformation.

\medskip

\no {\bf Proof of Lemma~\ref{Lmtr2od1}.}
Let $p,q\in L_{loc}^1(\R),\l\in\C$.
Assume that Eq.~\er{1b}
has a solution $\eta\in C^{[2]}(\R)$ and
$\eta(x)\ne 0$ for all $x\in\R$.
Let $y\in C^{[2]}(\R)$ be a solution to Eq.~\er{1b} and
let the function $f$ has the form \er{deff1}.
Then
$$
f={y'\eta-y\eta'\/\sqrt \eta},\qq
f'={y^{[2]}\eta-y\eta^{[2]}\/\sqrt \eta}-{f\eta'\/2\eta}.
$$
This yields
\[
\lb{f''}
\begin{aligned}
f''={(y^{[2]})'\eta-y(\eta^{[2]})'\/\sqrt \eta}+{y^{[2]}\eta'-y'\eta^{[2]}\/\sqrt \eta}
-{y^{[2]}\eta-y\eta^{[2]}\/\sqrt \eta}{\eta'\/2\eta}
-{f'\eta'\/2\eta}-{f\/2}\Big({\eta'\/\eta}\Big)'
\\
={(y^{[2]})'\eta-y(\eta^{[2]})'\/\sqrt \eta}+{y^{[2]}\eta'-y'\eta^{[2]}\/\sqrt \eta}
-{y^{[2]}\eta-y\eta^{[2]}\/\sqrt \eta}{\eta'\/\eta}
+{f\/4}\Big({\eta'\/\eta}\Big)^2-{f\/2}\lt({\eta''\/\eta}-\Big({\eta'\/\eta}\Big)^2\rt)
\\
={(y^{[2]})'\eta-y(\eta^{[2]})'\/\sqrt \eta}-f{\eta^{[2]}\/\eta}
+{f\/4}\Big({\eta'\/\eta}\Big)^2
-{f\/2}\lt({\eta^{[2]}-p\eta\/\eta}-\Big({\eta'\/\eta}\Big)^2\rt).
\end{aligned}
\]
The functions $y,\eta$ satisfy equation \er{1b}, which yields
$$
(y^{[2]})'\eta-y(\eta^{[2]})'=- p(y'\eta-y\eta')=-p\sqrt \eta f.
$$
Substituting this identity into \er{f''} we obtain \er{2oequd1}--\er{defxipsi}.
Moreover,
${\eta'\/\eta},{\eta^{[2]}\/\eta}-({\eta'\/\eta})^2\in AC_{loc}(\R)$.
The identities \er{da1+} and \er{defxipsi} give
$V+{p\/2}\in AC_{loc}(\R)$ and
$f,f',f''+{p\/2}f\in AC_{loc}(\R)$.~\BBox

\subsection{Analysis of the potential}

Assume that $p,q\in L_{loc}^1(\R)$.
In what follows, we will mainly deal with the solution
$\eta(x,\l)$ to Eq.~\er{1b}
that satisfies the following conditions:

a) $\eta(\cdot,\l)\in C^{[2]}(\R)$ for all
$\l\in\mG$, where $\mG\ss\C$ is a domain such that
$S_\d\ss\mG$ for some $\d\in(0,\pi)$,
$S_\d=\{\l\in\C:|\arg\l|<\d\}$,

b) $\eta(x,\cdot)$ is analytic in $\mG$,

c) $\eta(x,\l)\ne 0$ for all $(x,\l)\in\R\ts \mG$,

d) $\eta$ satisfies
\[
\lb{asmg}
\sup_{x\in\R}|e^{-zx}\eta(x,\l)-1|=o(1),\ \
\sup_{x\in\R}|z^{-1}e^{-zx}\eta'(x,\l)-1|=o(1),\ \
\sup_{x\in\R}|z^{-2}e^{-zx}\eta^{[2]}(x,\l)-1|=o(1),
\]
as $|\l|\to\iy,\l\in S_\d$.
For this case we introduce the function
\[
\lb{defxipsi}
\xi={\eta'\/\eta}-z.
\]
Then the definition \er{defxipsi} implies
\[
\lb{idxi'}
\xi'={\eta^{[2]}\/\eta}-\Big({\eta'\/\eta}\Big)^2-p,
\]
the definition \er{defxipsi} and the identity  \er{idxi'} give
$\xi,\xi'+p\in AC_{loc}(\R)$.
Moreover,
the definition \er{defxipsi} and the asymptotics \er{asmg}
yield
\[
\lb{asxi}
\sup_{x\in\R}|z^{-1}\xi(x,\l)|=o(1),\qq
\sup_{x\in\R}|z^{-2}(\xi'(x,\l)+p(x))|=o(1),
\]
as $|\l|\to\iy,\l\in S_\d$.
The definition \er{da1+} gives
\[
\lb{da1++}
V=E-2p-{3\/2}\xi'-{3\/4}(\xi+z)^2.
\]

\begin{lemma}
Let $p,q\in L_{loc}^1(\R)$, let $\l\in\mG$,
and let the function $\xi$
be defined by \er{defxipsi}.
 Then

i) The function $\xi$ satisfies the differential equation
\[
\lb{eqxi}
(\xi'+p)'+3 z(\xi'+ p)+3 z^2\xi+3\xi(\xi'+p)+3 z\xi^2
+\xi^3-p\xi- z p+q=0.
\]

ii) The vector-valued function $ X$ given by
\[
\lb{defmYmX}
 X=\ma X_1\\  X_2\am=\ma -\o^2\xi +{i\/\sqrt3 z}(\xi'+p)\\
-\o\xi -{i\/\sqrt3 z}(\xi'+p)\am,
\]
satisfies: $ X(\cdot,\l)\in AC_{loc}(\R)$ and
\[
\lb{eqyi1}
 X'=i\sqrt3 z\Omega X+  W+\cK[ X]\cI_-,
\]
where
\[
\lb{defmW}
\Omega=\ma \o&0\\0&-\o^2\am,\qq
 W={i\/\sqrt3 }\Big( p\cI_+-{q \/ z}\cI_-\Big),\qq
\cI_+=\ma\o\\-\o^2\am,\qq \cI_-=\ma 1\\-1\am,
\]
\[
\lb{defmA1mY}
\cK[ X]={i p\/\sqrt3 z}( X_1+ X_2)
+3( X_1+ X_2)(\o X_1 -\o^2 X_2)
-i\sqrt3 ( X_1+ X_2)^2
-{i\/\sqrt3 z}( X_1+ X_2)^3.
\]

iii) The potential $V(x,E)$ satisfies
\[
\lb{cVthY}
V=-{p\/2}-{3z\/2}( \o^2 X_1+\o X_2)
-{3\/4}( X_1+ X_2)^2,\qq
z=\Big({4\/3}E\Big)^{1\/2}.
\]
\end{lemma}

\no {\bf Proof.}
i) The definition \er{defxipsi} implies $\eta(x)=\eta(0)e^{zx+\int_0^x\xi(s) ds}$.
Substituting this identity into Eq.~\er{1b} we obtain \er{eqxi}.

ii) We rewrite Eq.~\er{eqxi} in the vector form
\[
\lb{eqxi1}
\mX'=A_0\mX+ B+A_1[\mX],
\]
on $\R$, where
$$
\mX=\ma\mX_1\\\mX_2\am=\ma\xi\\ \xi'+p\am\in  AC_{loc}(\R),\qq
A_0=\ma 0&1\\-3 z^2&-3 z\am,
$$
$$
B=\ma-p\\z p- q\am,\qq
A_1[\mX]=\ma 0\\ p\mX_1-3\mX_1\mX_2-3 z\mX_1^2-\mX_1^3\am.
$$
Introduce the matrix
$$
U=\ma 1&1\\i\sqrt3\o z &-i\sqrt3\o^2z\am,\qq
U^{-1}=\ma -\o^2 &{i\/\sqrt3 z}\\
-\o &-{i\/\sqrt3 z}\am.
$$
Multiplying Eq.~\er{eqxi1} by $U^{-1}$ and using
$$
U^{-1}A_0U=i\sqrt3 z\Omega,\qqq
 X=\ma -\o^2 &{i\/\sqrt3 z}\\-\o &-{i\/\sqrt3 z}\am\ma\xi\\ \xi'+p\am
=U^{-1}\mX,
$$
$$
 W={i\/\sqrt3 }\ma \o p-{1\/ z} q\\-\o^2 p+{1\/ z}q\am
=\ma (\o^2 +{i\/\sqrt3 }) p-{i\/\sqrt3 z} q\\(\o -{i\/\sqrt3 }) p+{i\/\sqrt3 z} q\am
=U^{-1}B,\qq
\cK[ X]\cI_-
=U^{-1}A_1[U X],
$$
we obtain \er{eqyi1}.

iii) The definition \er{da1+} implies
\[
\lb{cVtmX}
V=-{p\/2}-{3\/2}\mX_2-{3z\/2}\mX_1 -{3\/4}\mX_1^2.
\]
The definition \er{defmYmX} yields
$$
\mX_1= X_1+ X_2,\qq \mX_2=i\sqrt3 z(\o X_1-\o^2 X_2).
$$
Substituting these identities into \er{cVtmX}
and using the identities $1+i\sqrt3\o=\o^2,1-i\sqrt3\o^2=\o$
we obtain \er{cVthY}.~\BBox

\medskip

\no {\bf Remark.} If $\p\in\cH\os\cH$ and the solution $\eta$
is the Floquet solution, then $\x$, and therefore $V$,
is an 1-periodic function, and $\xi,\xi'+p,V+{p\/2}\in AC(\T)$.

\subsection{Solutions of the transpose equation.}
In the following lemma we prove that the numerator $y'g-yg'$
in the definition \er{deff1}
is a solution of the transpose equation \er{1btr} and
 study relations between solutions to Eqs.~\er{1b} and \er{1btr}.

\begin{lemma}
\lb{Lmsolconjeq}
Let $p,q\in L_{loc}^1(\R),\l\in\C$. Then

i) If $f,g\in C^{[2]}(\R)$ are two linearly independent solutions of
Eq.~\er{1b}, then $\tilde y=fg'-f'g$ is a solution of Eq.~\er{1btr}.

ii) Let $\tilde y\in C^{[2]}(\R)$ be a non-trivial solution of Eq.~\er{1btr},
let $g\in C^{[2]}(\R)$ be a solution of Eq.~\er{1b} such that $g(x)\ne 0, x\in\R$,
and let $f\in C^{[2]}(\R)$ satisfy
\[
\lb{wtyfg}
\tilde y=fg'-f'g.
\]
Then
\[
\lb{idhf'''}
(f''+pf)'+pf'+q f-\l f={(G,JY)\/g^2},
\]
where $\l=z^3,G=(g,g',g^{[2]})^\top,\tilde Y=(\tilde y,\tilde y',\tilde y^{[2]})^\top$,
\[
\lb{h0g20}
(G,J\tilde Y)=g\tilde y^{[2]}-g'\tilde y'+g^{[2]}\tilde y=\const.
\]

\end{lemma}

\no {\bf Proof.}
i) We have
$$
\tilde y=\det\ma f&g\\f'&g'\am,\qq
\tilde y'=\det\ma f&g\\f^{[2]}&g^{[2]}\am,
$$
$$
\tilde y^{[2]}=\det\ma f&g\\(f^{[2]})'&(g^{[2]})'\am
+\det\ma f'&g'\\f^{[2]}&g^{[2]}\am-p\tilde y
=\det\ma f'&g'\\f^{[2]}&g^{[2]}\am,
$$
$$
(\tilde y^{[2]})'=-p\det\ma f&g\\f^{[2]}&g^{[2]}\am
+\det\ma f'&g'\\(f^{[2]})'&(g^{[2]})'\am
=-p\tilde y'-(\l-q)\tilde y.
$$
Therefore, $\tilde y$ satisfies Eq.~\er{1btr}.

ii) The identity \er{wtyfg} gives
$$
f=g\s,\qq \s=\int_0^x{\tilde y(s)\/g^2(s)}ds.
$$
Then
$$
f'=g'\s+ g\s',\qq f^{[2]}=g^{[2]}\s+2g'\s'+g\s''.
$$
Using
$$
\s'={\tilde y\/g^2},\qq \s''={\tilde y'\/g^2}-{2\tilde y g'\/g^3},
$$
we obtain
$$
f^{[2]}=g^{[2]}\s+{\tilde y'\/g}.
$$
Then
$$
(f^{[2]})'=(g^{[2]})'\s+g^{[2]}\s'
+{\tilde y^{[2]}-p\tilde y\/g}-{\tilde y'g'\/g^2},
$$
which yields
$$
(f^{[2]})'+pf'+(q-\l)f=(g^{[2]}+p g)\s'
+{\tilde y^{[2]}-p\tilde y\/g}-{\tilde y'g'\/g^2}
={g^{[2]}\tilde y\/g^2}
+{\tilde y^{[2]}\/g}-{\tilde y'g'\/g^2}.
$$
This gives \er{idhf'''}.
The identities
$$
\begin{aligned}
-(\tilde y^{[2]})'-p\tilde y'+(q-\l)\tilde y=0,\\
(g^{[2]})'+pg'+(q-\l)g=0
\end{aligned}
$$
give
$$
(\tilde y^{[2]})'g+\tilde y(g^{[2]})'+p\tilde y'g+p\tilde yg'=0.
$$
This implies
$$
\big(\tilde y^{[2]}g+\tilde yg^{[2]}\big)'
-\tilde y'\big(g^{[2]}-pg\big)-\big(\tilde y^{[2]}-p\tilde y\big)g'=0.
$$
Then we have
$$
\big(\tilde y^{[2]}g+\tilde yg^{[2]}-\tilde y'g'\big)'=0,
$$
which yields \er{h0g20}.
\BBox

\section{Local results}
\setcounter{equation}{0}

\subsection{Transformation of the ramifications}

We consider the ramifications.

\medskip

\no {\bf Proof of Theorem~\ref{Lmtpevsch}.}
Assume, for definiteness, that $\t=\t_3$,
the proofs for $\t=\t_1$ and $\t=\t_2$ are similar.
 Assume that $\t(\l_o)$ for some $\l_o\in\C$ is a
simple multiplier and the corresponding Floquet solution
$\eta=\gf_3$ satisfies: $\eta(x,\l_o)\ne 0$ for all $x\in\R$.
Then the function $\t$ is analytic on some
neighborhood $\cU$ of $\l_o$ and $\eta(x,\cdot),x\in\R$, is also analytic
on $\cU$ and $\eta(x,\l)\ne 0$ for all $(x,\l)\in\R\ts\cU$,
see \cite{BK24xxx}.
Then each of the potentials $V(x,\cdot),x\in\R$, is analytic on $\cU$.

In this case $\l_o$ is a ramification for Eq.~\er{1b}
iff $\t_1(\l_o)=\t_2(\l_o)$.
Let $\gf_1,\gf_2$ be the corresponding Floquet solutions
defined by \er{Fs1}.
Then $f_1=\eta^{3\/2}({\gf_1\/\eta})'$
is the Floquet solution of Eq.~\er{2oequd1} satisfying
$f_1(x+1,E_o)=t_1(E_o)f_1(x,E_o)$,
where $t_1=\tau_3^{1\/2}\tau_1$ is a multiplier for  Eq.~\er{2oequd1},
$E_o={3\/4}\l_o^{2\/3}$.
Similarly, $f_2=\eta^{3\/2}({\gf_2\/\eta})'$
is the Floquet solution of Eq.~\er{2oequd1} satisfying
$f_2(x+1,E_o)=t_2(E_o)f_2(x,E_o)$,
where $t_2=\tau_3^{1\/2}\tau_2$ is the other multiplier for  Eq.~\er{2oequd1}.
We have $\t_1(\l_o)=\t_2(\l_o)$ iff $t_1(E_o)=t_2(E_o)$.
Therefore, $\l_o$ is a ramification for Eq.~\er{1b} iff
$E_o={3\/4}\l_o^{2\/3}$
is an eigenvalue of the 2-periodic problem for Eq.~\er{2oequd1}.~\BBox

\subsection{Monodromy matrices}
Introduce the monodromy matrix $M(\l)$ for Eq.~\er{1b} by \er{defmm}
and the corresponding monodromy matrix $\tilde M(\l)$
for Eq.~\er{1btr}.
They are entire $3\ts 3$ matrix-valued functions.
They satisfy
\[
\lb{symMwtM}
\tilde M^\top JM=J,\qq J=\ma 0&0&1\\0&-1&0\\1&0&0\am,
\]
see \cite{BK24xx}.
Recall that the matrix $M$ has three eigenvalues (counting with multiplicity)
$\t_1,\t_2,\t_3$.
The following lemma shows that
the matrix $\tilde M^\top$ (and then the matrix $\tilde M$) has three eigenvalues
$\t_1^{-1},\t_2^{-1},\t_3^{-1}$.

\begin{lemma}
\lb{LmevMtop}
Let $\p\in\cH\os\cH$. Then

i) $\t\in\C\sm\{0\}$ is an eigenvalue
of the matrix $M$ (and then $M^\top$) iff $\t^{-1}$ is an eigenvalue
of the matrix $\tilde M$.

ii) Let, in addition, $\cF$ be the eigenvector of the matrix $M^\top$
corresponding to the  eigenvalue $\t$.
Then $J\cF$ is the eigenvector of the matrix $\tilde M$
corresponding to the eigenvalue $\t^{-1}$,
here $J$ is given by \er{symMwtM}.
\end{lemma}

\no {\bf Proof.}
i) The identity \er{symMwtM} gives
$
M= J(\tilde M^\top)^{-1}J.
$
Let $\t\in\C\sm\{0\}$ be an eigenvalue
of the matrix $M$. The identities
$$
\begin{aligned}
\det(M-\t\1_3)=\det\big( J(\tilde M^\top)^{-1}J-\t\1_3\big)
=\det\big( (\tilde M^\top)^{-1}-\t\1_3\big)
\\
=-\t(\det\tilde M^\top)^{-1}\det(\tilde M^\top-\t^{-1}\1_3)
=-\t(\det\tilde M)^{-1}\det(\tilde M-\t^{-1}\1_3)
\end{aligned}
$$
give the statement.

ii)  Let $\t\in\C\sm\{0\}$ be an eigenvalue
of the matrix $M^\top$ and let  $\cF$ be the corresponding eigenvector.
The identity \er{symMwtM} implies
$M^\top =J\tilde M^{-1}J$.
Then the identity $M^\top\cF=\t\cF$
gives $J\tilde M^{-1}J\cF=\t\cF$, then we have $\tilde M^{-1}J\cF=\t J\cF$.
This yields $\tilde MJ\cF=\t^{-1}J\cF$, which proves the statement.~\BBox

\subsection{Monodromy matrix and three-point eigenvalues}

We prove the following results about the three-point problem.

\begin{lemma}
\lb{Lm3pev}
Let $\p\in\cH\os\cH$.

i) If the monodromy matrix $M(\l),\l\in\C$, has the eigenvector of the form
$(0,a,b)^\top,a,b\in\C$, then $\l$ is the eigenvalue of the three-point problem
for Eq.~\er{1b} and
$$
y=a\vp_2+b\vp_3
$$
is the corresponding eigenfunction.

ii) If $\l\in\C$ is the eigenvalue of the three-point problem
for Eq.~\er{1b}, then the corresponding eigenfunction has the form
\[
\lb{ef3p}
y(x)=\det\ma\vp_2(x,\l)&\vp_3(x,\l)\\\vp_2(2,\l)&\vp_3(2,\l)\am,\qq
x\in\R.
\]
Moreover,
\[
\lb{yx+1}
y(x+1)=Ay(x)
\]
for all $x\in\R$ and for some $A\in\C\sm\{0\}$, that is, $y$ is the Floquet
solution with the multiplier $A$. Moreover, the eigenvector of the monodromy
matrix $M$, corresponding the eigenvalue $A$, has the form
$(0,y'(0),y^{[2]}(0))^\top$.
\end{lemma}

\no {\bf Proof.}
i) Let the matrix $M(\l)$, has the eigenvector
$(0,a,b)^\top$. Then we have
$$
\ma y\\y'\\y^{[2]}\am(1)=M\ma y\\y'\\y^{[2]}\am(0)
=M\ma 0\\a\\b\am=A\ma 0\\a\\b\am=\ma 0\\aA\\bA\am,
$$
for some $A\in\C$, and
$$
\ma y\\y'\\y^{[2]}\am(2)=M^2\ma y\\y'\\y^{[2]}\am(0)
=M^2\ma 0\\a\\b\am=A^2\ma 0\\a\\b\am=\ma 0\\aA^2\\bA^2\am.
$$
This yields $y(0)=y(1)=y(2)=0$, that is,
$y=a\vp_2+b\vp_3$ is a solution of the three-point
problem.

ii) The function $y$, given by \er{ef3p}, satisfies Eq.~\er{1b}
and the three-point Dirichlet conditions. Moreover, due to the periodicity
of $p,q$, we have
\[
\lb{yx+1pr}
y(x+1)=A_1\vp_1(x)+A_2\vp_2(x)+A_3\vp_3(x)
\]
for some constant $A_1,A_2,A_3$.
The condition $y(1)=0$ gives $A_1=0$. The condition $y(2)=0$
implies
$$
A_2\vp_2(2)+A_3\vp_3(2)=0,
$$
which yields $A_3=-A_2{\vp_2(2)\/\vp_3(2)}$. Substituting these identities
into \er{yx+1pr} we obtain
$$
y(x+1)=A_2\Big(\vp_2(x)-{\vp_2(2)\/\vp_3(2)}\vp_3(x)\Big).
$$
This gives \er{yx+1}.

The Floquet solution $y$ satisfies
$$
y=y(0)\vp_1+y'(0)\vp_2+y^{[2]}(0)\vp_3.
$$
The identity \er{yx+1} yields
$$
M\ma y\\y'\\y^{[2]}\am(0)=\ma y\\y'\\y^{[2]}\am(1)
=A\ma y\\y'\\y^{[2]}\am(0),
$$
therefore, $(y(0),y'(0),y^{[2]}(0))^\top$ is the eigenvector of the monodromy
matrix $M$, corresponding the eigenvalue $A$. The condition $y(0)=0$
gives that this eigenvector has the form
$(0,y'(0),y^{[2]}(0))^\top$.~\BBox

\subsection{Additional results about the transpose equation}

If $y,g$ are solutions of Eq.~\er{1b},
then $\tilde y=y'g-yg'$ is a solution of
Eq.~\er{1btr}. We prove the following converse result.

\begin{lemma}
\lb{Lm3pef}
Let $\p\in\cH\os\cH$.
Let $\tilde y\in C^{[2]}(\R)$ be a solution of the three-point problem for
Eq.~\er{1btr} for some $\l\in\C$ and let there exist the Floquet solution
$g\in C^{[2]}(\R)$ of  Eq.~\er{1b} with the simple multiplier such that
$g$ does not vanish for all $x\in\R$ and
$g^{[2]}(0)\ne 0$.
Then there exists the solution $y\in C^{[2]}(\R)$ of Eq.~\er{1b} such that
$\tilde y=y'g-yg'$.
\end{lemma}

\no {\bf Proof.}
Let $\tilde y$ be a solution of the three-point problem for
Eq.~\er{1btr}.
Due to Lemma~\ref{Lmsolconjeq}~ii), there exists
the function $y$ such that
\[
\lb{wtyyf3}
\tilde y=yg'-y'g.
\]
Moreover, $y$ satisfies
\[
\lb{eqyC}
(y''+py)'+py'+q y-\l y={C\/g^2},
\]
where
\[
\lb{constCpr}
C=g\tilde y^{[2]}-g'\tilde y'+g^{[2]}\tilde y=\const.
\]
Furthermore, $\wt y$ is the Floquet solution, that is,
\[
\lb{wtyx+1}
\tilde y(x+1)=A\tilde y(x)
\]
for all $x\in\R$
and for some constant $A\ne 0$, see Lm~\ref{Lm3pev}. Then
$$
\tilde y(n)=0,\qq
\tilde y'(n)=A^n\tilde y'(0),\qq
\tilde y^{[2]}(n)=A^n\tilde y^{[2]}(0),\qq \forall\ n\in\Z.
$$
The identity
$g(x+1)=\t g(x)$, $\t$ is the multiplier for $g$, implies
$$
g(n)=\t^n g(0),\qq g'(n)=\t^n g'(0),\qq g^{[2]}(n)=\t^n g^{[2]}(0),\qq
\forall\ n\in\Z.
$$
The identity \er{constCpr} yields
\[
\lb{idCpr2}
C=\t^nA^n\big(g(0)\tilde y^{[2]}(0)-g'(0)\tilde y'(0)\big)=\const
\qq \forall\ n\in\Z.
\]
This yields $C=0$ or $A=\t^{-1}$.
If $A=\t^{-1}$, then the identity
\er{wtyx+1} gives that $\tilde y$ is the Floquet solution  of Eq.~\er{1btr}
with the multiplier $\t^{-1}$ and $\tilde y=\a \vp_2+\b\vp_3$
for some $\a,\b\in\R$. Then the Floquet solution $g$ of
Eq.~\er{1b} satisfying the multiplier $\t$ has the form
$g=\b\vp_1-\a\vp_2$, see Lm~\ref{LmevMtop}. This yields $g^{[2]}(0)=0$,
which contradicts the conditions of the lemma.
Thus, $C=0$
and the identities \er{eqyC} and \er{wtyyf3} show that $y$ is the solution
of Eq.~\er{1b} such that
$\tilde y=y'g-yg'$.~\BBox

\subsection{Transformation of the three-point eigenvalues}
We are ready to prove our first results about the three-point eigenvalues.

\medskip

\no {\bf Proof of Theorem~\ref{LmtpevschDir}.}
We have proved in Theorem~\ref{Lmtpevsch} that
if $\t(\l)$ is a simple multiplier for some
$\l\in\C$ and the Floquet solution
$\eta\in C^{[2]}(\R)$ of  Eq.~\er{1b} does not vanish for all $x\in\R$,
then
the potential $V(x,\cdot),x\in\R$, is analytic on
some neighborhood of $E$.

Let, in addition, $\l$ be an eigenvalue of the three-point Dirichlet problem
for Eq.~\er{1btr}, let $\tilde y$ be the corresponding eigenfunction, and let
$\eta^{[2]}(0)\ne 0$.
Due to Lemma~\ref{Lm3pef}, there exists
the solution $y$ of Eq.~\er{1b} such that
$\tilde y=y'\eta-y\eta'$. Then
$E={3\/4}\l^{2\/3}$ is an eigenvalue of the Dirichlet problem for
Eq.~\er{2oequd1} and $f={\tilde y\/\sqrt \eta}$
is the corresponding eigenfunction.

 Conversely, let $E$ be an eigenvalue of the Dirichlet problem for
Eq.~\er{2oequd1} and let $f$ be the corresponding eigenfunction.
Then $f(0)=f(1)=0$ and $f(x)=C\vp(x)$ for some constant $C$, where $\vt,\vp$ are the standard
solution $-y''+Vy=Ey,\vp(0)=\vt'(0)=0,\vt(0)=\vp'(0)=1$.
Using the periodicity $V$ we obtain
$\vp(x+1)=A\vt(x)+B\vp(x)$ for some constant $A,B$.
The condition $\vp(1)=A=0$ gives
$\vp(x+1)=B\vp(x)$, which yields $\vp(2)=B\vp(1)=0$, therefore,
$f(2)=0$.
Then the function $\tilde y=f\sqrt \eta$ is a solution of the
three-point Dirichlet problem for Eq.~\er{1btr} and $\l=({4\/3}E)^{3\/2}$
is the corresponding eigenvalue.~\BBox

\section{Estimates of the potential}
\setcounter{equation}{0}

\subsection{Preliminaries}
We prove that the spectrum of the Schr\"odinger operator
with an energy-dependent potential is real using the results of our paper
\cite{BK21}.
For this we need estimates of the potential.

Introduce the diagonal matrix
\[
\lb{defcE}
\cE=\big(\1_2-e^{i\sqrt3 z\Omega}\big)^{-1}e^{i\sqrt3 z\Omega t}
=\ma\cE_1&0\\0&\cE_2\am,\qq
\cE_1= {e^{i\sqrt3 z\o t}\/1-e^{i\sqrt3 z\o}},
\qq \cE_2={e^{-i\sqrt3 z\o^2 t}\/1-e^{-i\sqrt3 z\o^2}},
\]
we used the definition \er{defmW} of $\Omega$. Moreover,
\[
\lb{det1-expA0}
\big(\1_2-e^{i\sqrt3 z\Omega}\big)^{-1}
={\1_2-e^{-i\sqrt3 z\Omega^*}\/D},
\]
where
\[
\lb{defD1}
\Omega^*=\ma \o^2&0\\0&-\o\am,\qq
D(\l)=(1-e^{i\sqrt3 z\o})(1-e^{-i\sqrt3 z\o^2})
=4\sin{\sqrt3\o z\/2}\sin{\sqrt3\o^2 z\/2}.
\]
The zeros of the function $D$ in $\C\sm\{0\}$ are simple and have the form
$-({2\pi n\/\sqrt3})^3,n\in\N$.
We need the following simple estimates.

\begin{lemma}
i) Let $(x,\l)\in\R\ts\C_+,f\in L^2(\T)$. Then
\[
\lb{estintOA}
\Big|\int_0^1e^{i\sqrt3 z\Omega t}f(x-t)ds\Big|\le 2\|f\|.
\]

ii) If $\l\in\mD_3$, the domain $\mD_3$ is given by \er{defmD3},
then the following estimates hold true:
\[
\lb{lowestexp}
|1-e^{i\sqrt3 \o z}|\ge{1\/2},\qq |1-e^{-i\sqrt3 \o^2 z}|\ge{1\/2},
\]
\[
\lb{lowestD}
|D(\l)|\ge{1\/4}.
\]
Moreover,
\[
\lb{estinvIO}
\Big|\big(\1_2-e^{i\sqrt3 z\Omega}\big)^{-1}\Big|\le 16.
\]

\end{lemma}

\no {\bf Proof.}
i) The definition \er{defmW} of $\Omega$ implies
$$
|e^{i\sqrt3 z\Omega x}|\le|e^{i\sqrt3 z\o x}|+|e^{-i\sqrt3 z\o^2 x}|
=2e^{-{3\/2}x\Re z}\cosh\Big({\sqrt3\/2}x\Im z\Big),\qqq (x,\l)\in\R_+\ts\C.
$$
We have $-\sqrt3\Re z+\Im z\le 0$ for all $\arg z\in[0,{\pi\/3}]$.
Then $e^{-{3\/2}x\Re z}\cosh({\sqrt3\/2}x\Im z)<1$,
which yields \er{estintOA}.

ii) Note that $\Im (\o z)\ge 0,\Im (\o^2 z)\le 0$,
for all $\l\in\C$. Let $\l\in\mD_3$. Then
$$
|1-e^{i\sqrt3 \o z}|=2|e^{i{\sqrt3\o z\/2}}|\Big|\sin{\sqrt3\o z\/2}\Big|
\ge{1\/2}e^{-{\sqrt 3\/2}\Im (\o z)}e^{{\sqrt 3\/2}|\Im (\o z)|}={1\/2},
$$
$$
|1-e^{-i\sqrt3 \o^2z}|=2|e^{-i{\sqrt3\o^2 z\/2}}|\Big|\sin{\sqrt3\o^2 z\/2}\Big|
\ge{1\/2}e^{{\sqrt 3\/2}\Im (\o^2 z)}e^{{\sqrt 3\/2}|\Im (\o^2 z)|}={1\/2},
$$
where we used the estimates
\[
\lb{lestsinoo2}
\Big|\sin{\sqrt 3 \o z\/2}\Big|\ge{1\/4}e^{{\sqrt 3\/2}|\Im \o z|},\qq
\Big|\sin{\sqrt 3 \o^2 z\/2}\Big|\ge{1\/4}e^{{\sqrt 3\/2}|\Im \o^2 z|},\qq
\l\in\mD_3,
\]
see \cite[Lm~2.1]{PT87}. This yields \er{lowestexp}.
The definition \er{defD1} yields \er{lowestD}.
The estimate \er{lowestD}
and the identity \er{det1-expA0} give
$$
\Big|\big(\1_2-e^{i\sqrt3 z\Omega}\big)^{-1}\Big|\le 4
\big|\1_2-e^{-i\sqrt3 z\Omega^*}\big|\le 4\big(|1-e^{-i\sqrt3 z\o^2}|
+|1-e^{i\sqrt3 z\o}|\big)\le 16,
$$
which yields \er{estinvIO}.~\BBox

\subsection{Transformation to an integral equation in the periodic case}
In the case of  periodic coefficients $p,q$ we
rewrite Eq.~\er{eqyi1} into the form of the integral equation \er{eqwtmXG}.

\begin{lemma}
Let $(\l,\p)\in\C\sm\{({2\pi n\/\sqrt3})^3,n\in\Z\}\ts\cH\os\cH$.
Let $ X(\cdot,\l)\in AC(\T)$ be an 1-periodic solution of
Eq.~\er{eqxi1}. Then $ X$ satisfies the non-linear integral equation
\[
\lb{eqwtmXG}
 X=G[ X],
\]
where
\[
\lb{defmapG}
G[ X]= X^0+F[ X],
\]
\[
\lb{defmX01}
 X^0(x,\l)=
{i\/\sqrt3 }
\int_0^1 \cE(t,\l)\Big( p(x-t)\cI_+-{q(x-t)\/ z}\cI_-\Big)dt
\]
($\cE,\cI_\pm$ are given by \er{defcE} and \er{defmW})
is the 1-periodic solution of the equation
\[
\lb{eqxi1mX0}
( X^0)'=i\sqrt3 z\Omega X^0+  W,
\]
and the integral operator $F$ is given by
\[
\lb{defFmX}
F[ X](x,\l)=
\int_0^1  \cE(t,\l)\cK[ X](x-t,\l)\cI_-dt,
\]

\end{lemma}

\no {\bf Proof.}
Let $\l\in\mD_3$, $f\in L^1(\T)$. Then the 1-periodic
solution $ X\in AC(\T)$ of the equation
\[
\lb{eqmXF}
 X'=i\sqrt3 z\Omega X+ f
\]
has the form
\[
\lb{soleqmXF}
 X(x,\l)=\int_{x-1}^x \cE(x-t,\l)f(t)dt
=\int_0^1  \cE(t,\l)f(x-t)dt.
\]
Put $f= W+\cK[ X]\cI_-$ in \er{eqmXF}, then we conclude that
the 1-periodic solution of Eq.~\er{eqxi1} satisfies
the integral equation \er{eqmX},
\[
\lb{eqmX}
 X= X^0+F[ X],
\]
and then \er{eqwtmXG}, where
$$
 X^0(x,\l)=
\int_0^1 \cE(t,\l) W(x-t,\l)dt.
$$
satisfies Eq.~\er{eqxi1mX0},
and the integral operator $F$ is given by \er{defFmX}.
The definition \er{defmW} gives
\er{defmX01}.~\BBox

\subsection{Solvability of Eq.~\er{eqwtmXG} }
Introduce the spaces $C(\T)$ of continuous scalar and
2-vector  functions on the circle
equipped with the norm
$$
\|f\|_\iy=\max_{x\in\T}|f(x)|.
$$
Introduce the balls
$$
\mB(\d)=\{\cA\in C(\T):\|\cA\|_\iy<\d\}\ss C(\T),\qq
\d>0,
$$
of 2-vector  functions on the circle.

We consider the mapping $G:C(\T)\to C(\T)$ given by \er{defmapG}.
We will show that there exists $\d_0>0$ such that the mapping $G$
is a contraction on the ball $\mB(\d_0)$. Then Eq.~\er{eqwtmXG}
has a unique solution in this ball. In order to obtain this solution
we construct the sequence $( X^n)_{n=0}^\iy\ss\mB(\d_0)$ of iterations.
We take $ X^0$ as a first approximation and then the next
approximations $ X^n,n\in\N$, are given by
$ X^{n}=G[ X^{n-1}]$ for all $n\in\N$.
Then the solution $ X$ of equation \er{eqwtmXG} has the form
$$
 X=\lim_{n\to\iy} X^n.
$$
Moreover,
$$
 X^n= X^0+\sum_{k=1}^n( X^k- X^{k-1}),
$$
which yields
\[
\lb{sumwtmX}
 X= X^0+\sum_{k=1}^\iy( X^k- X^{k-1}).
\]

\begin{lemma}
\lb{LmmYpq}
i)  Let $(\l,\p)\in\mD_3\ts\cH\os\cH$ and let 2-vector-valued
functions $ X, Y\in C(\T)$. Then
\[
\lb{GmZ0mZ}
\big\|G[ X]-G[ Y]\big\|_\iy
\le{64\/\sqrt3 }\| X- Y\|_\iy
\Big(\|p\|+9\big(\| X\|_\iy+\| Y\|_\iy\big)
+\big(\| X\|_\iy+\| Y\|_\iy\big)^2\Big).
\]

ii) Let $(\l,\p)\in\mD_3\ts\cB_\C(\ve_0)$.
If, in addition, $\max\{\| X\|_\iy,\| Y\|_\iy\}\le\d_0$
for some $\d_0>0$ small enough, then
\[
\lb{GmY-mZ45}
\big\|G[ X]-G[ Y]\big\|_\iy
\le{1\/2}\| X- Y\|_\iy.
\]

iii)  Let $(\l,\p)\in\mD_3\ts\cB_\C(\ve_0)$
and let $ X^0(\cdot,\l)\in\mB(\d_0)$.
Then Eq.~\er{eqwtmXG} has the unique solution $ X$ in the ball $\mB(\d_0)$.
Moreover, this solution satisfies
\[
\lb{estmY}
\| X(\cdot,\l)\|_\iy
\le C\| X^0(\cdot,\l)\|_\iy,
\]
for some $C>0$.
Each of the functions $ X(\cdot,\cdot,\p),\p\in\cB_\C(\ve_0)$, is analytic
on the domain $\mD_3$.
Each of the functions $ X(\cdot,\l,\cdot),\l\in\mD_3$, is analytic on the ball
$\cB_\C(\ve_0)$.
\end{lemma}

\no {\bf Proof.}
i) Let $\l\in\mD_3$. The definition \er{defFmX} and the estimate
\er{estinvIO} imply
\[
\lb{estFmY}
\big|F[ X](x)\big|\le 32\|\cK[ X]\cI_-\|
\]
for all $x\in\T$ and
for a 2-vector-valued function $ X\in C(\T)$,
here and below in this proof
$F[ X](x)=F[ X](x,\l), X^0(x)= X^0(x,\l),...$
Then the definition \er{defmapG} gives
\[
\lb{estmY1-mY0}
\big| X^1(x)- X^0(x)\big|=\big|G[ X^0](x)- X^0(x)\big|=\big|F[ X^0](x)\big|
\le 32\|\cK[ X^0]\cI_-\|,
\]
for all $x\in\T$.
Moreover,
\[
\lb{estG-Gpr}
\big|G[ X](x)-G[ Y](x)\big|
=\big|F[ X](x)-F[ Y](x)\big|
\le32\|\cK[ X]\cI_--\cK[ Y]\cI_-\|
\]
for two 2-vector-valued functions $ X, Y\in C(\T)$.
The definitions \er{defmW} and \er{defmA1mY} give
\[
\lb{mA=2mA1}
\|\cK[ X]\cI_-\|=2\|\cK[ X]\|,\qq
\big\|\cK[ X]\cI_--\cK[ Y]\cI_-\big\|=2\big\|\cK[ X]-\cK[ Y]\big\|.
\]
The definition \er{defmA1mY} and the estimate $|z|>1$ imply
\[
\lb{estmA0}
\|\cK[ X]\|
\le{\| X\|_\iy\/\sqrt3 }\big( \|p\|+9\| X^0\|_\iy+\| X\|_\iy^2\big).
\]
Moreover,
$$
\begin{aligned}
\cK[ X]-\cK[ Y]=
{i p\/\sqrt3 z}( X_1- Y_1+ X_2- Y_2)
+3( X_1+ X_2)\big(\o( X_1- Y_1) +\o^2( Y_2 - X_2)\big)
\\
+3( X_1- Y_1+ X_2- Y_2)(\o Y_1 -\o^2 Y_2)
+i\sqrt3 \big(( Y_1+ Y_2)^2-( X_1+ X_2)^2\big)
\\
+{i\/\sqrt3 z}\big(( Y_1+ Y_2)^3-( X_1+ X_2)^3\big),
\end{aligned}
$$
which yields
$$
\Big\|\cK[ X]-\cK[ Y]\Big\|
\le{\| X- Y\|_\iy\/\sqrt3 }
\Big( \|p\|
+9\big(\| X\|_\iy+\| Y\|_\iy\big)
+\big(\| X\|_\iy+\| Y\|_\iy\big)^2\Big).
$$
Then the estimate \er{estG-Gpr} and the identities \er{mA=2mA1} give
 \er{GmZ0mZ}.

ii) Let $(\l,\p)\in\mD_3\ts\cB_\C(\ve_0)$. Then $|z|>1$ and
$\|p\|\le \ve_0$.
If $\max\{\| X\|_\iy,\| Y\|_\iy\}\le\d_0$,
then
the estimate \er{GmZ0mZ} yields \er{GmY-mZ45}.

iii) Due to \er{GmY-mZ45}, the mapping $G$ is a contraction on the ball
$\mB(\d_0)\ss C(\T)$.
Then equation \er{eqwtmXG} has the unique
solution $ X$ in this ball.
The estimate \er{GmY-mZ45} implies
$$
\| X^k- X^{k-1}\|_\iy=\|G[ X^{k-1}]-G[ X^{k-2}]\|_\iy
\le{1\/2}\| X^{k-1}- X^{k-2}\|_\iy\le
{1\/2^{k-1}}\| X^1- X^0\|_\iy.
$$
This shows that the series \er{sumwtmX} converges absolutely and uniformly
on any bounded subset of $\mD_3\ts\cB_\C(\ve_0)$.
Each term is an analytic function on $\mD_3$ for fixed
$\p\in\cB_\C(\ve_0)$, and  on $\cB_\C(\ve_0)$ for fixed
 $\l\in\mD_3$. Then the sum is also an analytic function.
 Summing the majorants we obtain
$$
\| X- X^0\|_\iy\le
\| X^1- X^0\|_\iy\Big(1+\sum_{k=2}^\iy{1\/2^{k-1}}\Big)
=2\| X^1- X^0\|_\iy\le 64\|\cK[ X^0]\cI_-\|\le 128\|\cK[ X^0]\|,
$$
where we used \er{estmY1-mY0} and  \er{mA=2mA1}. The estimate
\er{estmA0} gives
$$
\| X(\cdot,\l)- X^0(\cdot,\l)\|_\iy
\le C\| X^0(\cdot,\l)\|_\iy,
$$
for some $C>0$.
Using the estimate
$$
\| X\|_\iy\le \| X^0\|_\iy+\| X- X^0\|_\iy,
$$
we obtain \er{estmY}.
\BBox

\subsection{Estimates in the class $p',q\in L^2(\T)$}
We have the following estimate of $ X_0$ and $ X$ in the case
 $p',q\in L^2(\T)$.

\begin{lemma}
\lb{LmestmYpot}
Let $\p\in\cH_{1,\C}\os\cH_\C$ and let $\l\in\mD_3$. Then

i) The function $ X^0$, given by  \er{defmX01}, satisfies
\[
\lb{estY0'}
\big\| X^0(\cdot,\l)\big\|_\iy\le{19\|\p\|_1\/|z|}.
\]

ii) The function $F[ X^0]$, given by \er{defFmX}, satisfies
\[
\lb{estFmY01}
\|F[ X^0](x,\l)\|_\iy\le{C\|\p\|_1^2\/|z|^3},
\]
for some $C>0$.

iii) Each of the functions
$ X(x,\cdot,\cdot),x\in\T$,
is analytic with respect to $\l$ on $\mD_3$ for fixed
$\p\in\cB_{1,C}(\ve_0)$, and it is analytic with respect to $\p$
on $\cB_{1,C}(\ve_0)$ for fixed $\l\in\mD_3$.
Moreover, if $(\l,\p)\in\mD_3\ts\cB_{1,\C}(\ve_0)$,
then
\[
\lb{estmYi}
\| X(\cdot,\l)\|_\iy
\le{C\|\p\|_1\/|z|},
\]
\[
\lb{estmY-mY0imp}
\| X(\cdot,\l)- X^0(\cdot,\l)\|_\iy\le{C\|\p\|_1^2\/|z|^3},
\]
for some $C>0$.

\end{lemma}

\no {\bf Proof.}
i) We have
$$
\begin{aligned}
\int_0^1 e^{i\sqrt3 z\o t}p(x-t)dt={1\/i\sqrt3 z\o}\Big(
(e^{i\sqrt3 z\o}-1)p(x)
+\int_0^1 e^{i\sqrt3 z\o t}p'(x-t)dt\Big),
\\
\int_0^1 e^{-i\sqrt3 z\o^2 t}p(x-t)dt=-{1\/i\sqrt3 z\o^2}\Big(
(e^{-i\sqrt3 z\o^2}-1)p(x)
+\int_0^1 e^{-i\sqrt3 z\o^2 t}p'(x-t)dt\Big).
\end{aligned}
$$
This yields
$$
\int_0^1 e^{i\sqrt3 z\Omega t}p(x-t)dt=
{\Omega^{-1}\/i\sqrt3 z}\Big(
(e^{i\sqrt3 z\Omega}-\1_2)p(x)
+\int_0^1 e^{i\sqrt3 z\Omega t}p'(x-t)dt\Big).
$$
The definition \er{defmX01} and the identity
$\Omega^{-1}\cI_+=(1,1)^\top$, see \er{defmW}, imply
\[
\lb{defmX0}
 X^0(x,\l)={1\/3 z}\Big(-p(x)\ma 1\\1\am
+\int_0^1 \cE(t,\l)\big(p'(x-t)\ma 1\\1\am-i\sqrt3 q(x-t)\cI_-\big) dt\Big),
\]
for all $x\in\R$, where $\cE$ is given by \er{defcE}.
The identity \er{defmX0} and
the estimates \er{estintOA} and \er{estinvIO} give
$$
\big\| X^0(\cdot,\l)\big\|_\iy\le
{1\/3|z|}\Big(\|p\|_\iy+32(\|p'\|+\sqrt3 \|q\|)\Big)\le
{1\/3|z|}\Big(\|p\|+33\|p'\|+32\sqrt3 \|q\|\Big).
$$
here we used the estimate $\|p\|_\iy\le\|p\|+\|p'\|$.
This yields \er{estY0'}.

ii) The definitions \er{defFmX} and \er{defmW} imply
\[
\lb{estFmY0}
F[ X^0](x,\l)=
\int_0^1 \cE(t,\l)\cI_-\cK[ X^0](x-t,\l)dt
=
\int_{x-1}^x \cE(x-y,\l)\cI_-\cK[ X^0](y,\l)dy,
\]
where
$$
\cE\cI_-=(\cE_1,-\cE_2)^\top=\lt( {e^{i\sqrt3 z\o t}\/1-e^{i\sqrt3 z\o}},
-{e^{-i\sqrt3 z\o^2 t}\/1-e^{-i\sqrt3 z\o^2}} \rt)^\top,
$$
$\cE,\cE_1$, and $\cE_2$ are given by \er{defcE} and we used
$\cI_-=(1,-1)^\top$, see \er{defmW}.
The definition \er{defmA1mY} implies
\[
\lb{mA1mY0}
\cK[ X^0]={i p\/\sqrt3 z}( X_1^0+ X_2^0)
+3( X_1^0+ X_2^0)(\o X_1^0 -\o^2 X_2^0)
-i\sqrt3 ( X_1^0+ X_2^0)^2
-{i\/\sqrt3 z}( X_1^0+ X_2^0)^3.
\]
The identity \er{defmX0} gives
$$
\a X_1^0+\b X_2^0={(\b-\a)p+A_{\a\b}\/3 z},
$$
for all $\a,\b\in\C$, where
$$
\begin{aligned}
A_{\a\b}(x)=\int_0^1\Big( \big(\a\cE_1(t,\l)+\b\cE_2(t,\l)\big)p'(x-t)
-i\sqrt3 \big(\a\cE_1(t,\l)-\b\cE_2(t,\l)\big)q(x-t)\Big) dt
\\
=\int_{x-1}^{x}\Big( \big(\a\cE_1(x-y,\l)+\b\cE_2(x-y,\l)\big)p'(y)
-i\sqrt3 \big(\a\cE_1(x-y,\l)-\b\cE_2(x-y,\l)\big)q(y)\Big) dy,
\end{aligned}
$$
therefore, $A_{\a\b}'\in L^1(0,1)$.
The identity \er{mA1mY0} yields
$$
\cK[ X^0]={iA_{11}\/3\sqrt3 z^2}\Big(4 p-i\sqrt3A_{\o,-\o^2}
-A_{11}-{A_{11}^2\/9z^2}\Big).
$$
Integrating by parts in \er{estFmY0} we obtain \er{estFmY01}.

iii) The series \er{sumwtmX} converges absolutely and uniformly
on any bounded subset of $\mD_3\ts\cB_{1,C}(\ve_0)$.
Each term is an analytic function on $\mD_3$ for fixed
$\p\in\cB_{1,C}(\ve_0)$ and  on $\cB_{1,C}(\ve_0)$ for fixed
$\l\in\mD_3$. Then the sum is also analytic.
The estimates \er{estmY} and \er{estY0'} give
\er{estmYi}.

Eq.~\er{eqwtmXG} gives
\[
\lb{eqmYmY0F}
\| X- X^0\|_\iy=\|F[ X]\|_\iy\le \|F[ X]-F[ X^0]\|_\iy+\|F[ X^0]\|_\iy.
\]
The definitions \er{defFmX} and \er{defmW} imply
\[
\lb{estFmY0+FmY}
F[ X](x)-F[ X^0](x)=
\int_{x-1}^x \cE(x-y)\cI_-\Big(\cK[ X](y)-\cK[ X^0](y)\Big)dy,
\]
we omit $\l$ in the arguments.
The definition \er{defmA1mY} yields
$$
\begin{aligned}
\cK[ X]-\cK[ X^0]\!=\!{i p(F_1[ X]+F_2[ X])\/\sqrt3 z}
\!+\!3( X_1+ X_2)(\o X_1-\o^2 X_2)
\!-\!3( X_1^0+ X_2^0)(\o X_1^0-\o^2 X_2^0)
\\
-i\sqrt 3( X_1+ X_2)^2+i\sqrt 3( X_1^0+ X_2^0)^2
-{i( X_1+ X_2)^3\/\sqrt3 z}+{i( X_1^0+ X_2^0)^3\/\sqrt3 z},
\end{aligned}
$$
therefore,
$(\cK[ X]-\cK[ X^0])'\in L^1(0,1)$.
Integration by parts in  \er{estFmY0+FmY} and the estimates \er{estY0'}
and \er{estmYi} yield
$$
\big\|F[ X]-F[ X^0]\big\|_{\iy}
\le{C\|\p\|_1^2\/|z|^3},
$$
for some $C>0$. Substituting this estimate and \er{estFmY01} into \er{eqmYmY0F}
we obtain \er{estmY-mY0imp}.
\BBox

\medskip

Due to Lemma~\ref{LmestmYpot}~ii),
each function $ X(x,\cdot,\cdot),x\in\T$, is analytic on $\mD_3$ for fixed
$\p\in\cB_{1,\C}(\ve_0)$, and on $\cB_{1,\C}(\ve_0)$ for fixed
$\l\in\mD_3$. Then the identity \er{cVthY} shows that
each of the potentials
$V(x,\cdot,\cdot),x\in\T$,
is analytic with respect to $E$ on $\cS$ for fixed
$\p\in\cB_{1,\C}(\ve_0)$, and it is analytic with respect to $\p$
on $\cB_{1,\C}(\ve_0)$ for fixed $E\in\cS$.

\section{Second order equation.  Proof of Theorem~\ref{ThMcKtr}}
\setcounter{equation}{0}

\subsection{The spectra are real}
Each of the potential $V(x,\cdot),x\in\R$, is analytic on the domain
$\cS$, then there are well defined spectra of the 2-periodic problem
\er{fbc}, the Dirichlet problem \er{dbc}, and the quasi-periodic problems
\[
\lb{qpbc}
f(1)=e^{ik}f(0),\qqq f'(1)=e^{ik}f'(0),\qqq k\in[0,2\pi),
\]
for Eq.~\er{2oequd1} in this domain, see \cite{BK21}.
Moreover, these spectra are pure discrete.

Introduce the half-planes
\[
\lb{defhp}
\Pi_r=\{E\in\C:\Re E>r\},\qq r\in\R.
\]
The definition \er{mDBe} gives
\[
\lb{PiiimD}
\Pi_{3\/4}\ss\cS.
\]
where the half-planes $\Pi_r,r\in\R$, are defined by \er{defhp}.

\begin{lemma}
\lb{Lmrealsp}

i) The potential $V$ satisfies
\[
\lb{estcVr}
\Big\|V(x,E)+{p\/2}\Big\|_\iy\le {C\|\p\|_1},
\]
for all $(E,\p)\in\cS\ts\cB_{1,\C}(\ve_0)$ and  for some $C>0$.

ii) Let  $\p\in\cB_{1}(\ve_0)$. Then the spectra
of the problems \er{fbc}, \er{dbc}, and \er{qpbc}
for Eq.~\er{2oequd1}
in the half-plane $\Pi_2$ are all real.
\end{lemma}

\no {\bf Proof.}
i)The identity \er{cVthY} and the estimate \er{estmYi} imply \er{estcVr}.

ii) Recall the following result, see \cite[Th~1.3]{BK21}:

\no {\it
Let for some $a\in\R$ the potential $ V$ satisfies:

1) For each $E\in \Pi_a$ the function $ V$ is 1-periodic and
$ V(\cdot,E)\in L^1(\T)$,

2) For almost every $x\in\T$ the function $ V(x,\cdot)$ is real
analytic on the domain $\Pi_a$.

3) $\xi=\sup_{E\in\Pi_a}\|\Im V(x,E)\|_\iy<\iy.$

\no Then the spectra in the half-plane $\Pi_{a+r}$ are real:
$$
\gS\cap\Pi_{a+ r}\ss(a+ r,+\iy),
\qq\text{where}\qq
r={\xi\/2-\sqrt3}.
$$
}

The estimate \er{estcVr} gives
$$
\|\Im V(x,E)\|_\iy\le{C\|\p\|_1},\qq
\forall\qq (x,E)\in\T\ts\cS.
$$
Due to \er{PiiimD}, we may take
$a={3\/4}$ and $\xi=C\|\p\|_1$ in the above result.
Then $a+r<2$ and the spectra in the half-plane $\Pi_2$ are real.
\BBox

\subsection{Spectra}

Let  $\p\in\cB_\C(\ve_0)$.
Introduce  the fundamental solutions $\vt(x,E)$, $\vp(x,E),(x,E)\in\R\ts\cS$
of Eq.~\er{2oequd1} satisfying the initial
conditions $\vt(0,E)=\vp'(0,E)=1,\vt'(0,E)=\vp(0,E)=0$.
Introduce the Lyapunov function by
\[
\lb{fl2ovv}
\D( E)={1\/2}\big(\vt(1, E)+\vp'(1, E)\big),
\]
Each of the functions $\vt(x,\cdot)$, $\vp(x,\cdot)$,
$\vt'(x,\cdot)$, $\vp'(x,\cdot)$, $x\in\R$,
and then $\D$, is analytic on $\cS$, see \cite{BK21}.

The spectrum of the 2-periodic problem \er{2oequd1}, \er{fbc}
is the set of zeros of the functions
$\D(E)\pm1$ in the domain $\cS$,
the spectrum of the quasi-periodic problem \er{2oequd1}, \er{qpbc}
for $k\in[0,2\pi)$
is the set of zeros of the functions $\D(E)-\cos k$ in $\cS$,
the spectrum of the Dirichlet problem  \er{2oequd1}, \er{dbc}
is the set of zeros of the function
$\vp(1,E)$ in $\cS$.

In the unperturbed case $\p=0$
the spectrum of the 2-periodic problem
consists of the simple eigenvalue $E_0^{o,+}=0$ and the eigenvalues
$ E_{n}^{o,\pm}=(\pi n)^2, n\in\N$, of multiplicity 2,
the spectrum of the Dirichlet problem  \er{2oequd1}, \er{dbc}
consists of the simple eigenvalues $\gm_n^o=(\pi n)^2,n\in\N$.

We prove the following results.

\begin{lemma}
\lb{Lmalt2o}
Let $\p\in\cB_{1,\C}(\ve_0)$. Then

i) There are exactly two (counting with multiplicity)
eigenvalues $E_n^\pm$ of the 2-periodic problem \er{2oequd1}, \er{fbc}
in each domain $\cS_n,n\in\N$, and there are no eigenvalues
 in the domain $\cS\sm\cup_{n\in\N}\cS_{n}$. If $\p$ is real, then
 all zeros
$E_n^\pm,n\in\N$, are real,
may be labeled by \er{labEjpm} and satisfy
\[
\lb{Ejpm}
\begin{aligned}
\D(E_n^\pm)=(-1)^n,\qq (-1)^n\D(E)>1\ \ \forall\ E\in(E_n^-,E_n^+),\qq
n\ge 1,\\
-1<\D(E)<1,\qq\forall\ \
E\in(2,E_1^-)\cup\big(\cup_{n\ge 2}(E_{n-1}^+,E_n^-)\big),\\
\D'(E)<1\ \ \forall\
E\in(2,E_1^-)\cup\big(\cup_{n\ge 2}(E_{2n-1}^+,E_{2n}^-)\big),\qq
\D'(E)>1\ \ \forall\
E\in\cup_{n\ge 1}(E_{2n}^+,E_{2n+1}^-).
\end{aligned}
\]

ii) There are exactly one simple
eigenvalue $\gm_n$ of the Dirichlet problem  \er{2oequd1}, \er{dbc}
in each domain $\cS_n,n\in\N$, and there are no eigenvalues
 in the domain $\cS\sm\cup_{n\in\N}\cS_{n}$. If $\p$ is real,
 then these eigenvalues are real and
satisfy \er{gmj}.

\end{lemma}

\no {\bf Proof.}
i) The function $\D$ is analytic on $\cS$.
Moreover,
we have the estimate
$$
\big|\D(E)-\cos \sqrt E\big|
\!=\!\big|\D(E)-1+2\sin^2{\sqrt E\/2}\big|
\!=\!\big|\D(E)+1-2\cos^2{\sqrt E\/2}\big|
\!\le\! \| V(\cdot,E)\|e^{\| V(\cdot,E)\|+|\Im\sqrt E|},
$$
for all $ E\in\cS$,
see \cite[Lm~3.2~ii)]{BK21}.

Let $E\in\cS\sm\cup_{n\in\N}\cS_{2n}$. The definition \er{defOmegan}
gives $|{\sqrt E\/2}-\pi n|\ge{\sqrt 3\/4}$ for all $n\in\N$. Then
$$
|e^{|\Im{\sqrt E\/2}|}|<C|\sin{\sqrt E\/2}|,
$$
for some $C>0$, see \cite{PT87}.
The estimate \er{estcVr} gives
$$
\big|\D(E)-1+2\sin^2{\sqrt E\/2}\big|
\le C\|\p\|_1e^{C\|\p\|_1}e^{|\Im\sqrt E|}
< 2|\sin^2{\sqrt E\/2}|.
$$
Then, by Rouche's theorem, the functions $\D(E)-1$ and
$\sin^2{\sqrt E\/2}$ has the same number of zeros in each domain
$\cS_{2n},n\in\N$.
Since the function $\sin^2{\sqrt E\/2}$ has exactly one zero
of multiplicity two
at each point $2\pi n,n\in\Z$, and has no other zeros, the function
$\D(E)-1$ has exactly two zeros in each domain
$\cS_{2n},n\in\N$.
Moreover,
 we have the estimates
$$
\big|\D(E)-1\big|\ge \big|2\sin^2{\sqrt E\/2}\big|-
\big|\D(E)-1+2\sin^2{\sqrt E\/2}\big|>0,
$$
which shows that the function
$\D(E)-1$ has no zeros in the domain  $\cS\sm\cup_{n\in\N}\cS_{2n}$.

Similarly, if  $E\in\cS\sm\cup_{n\in\N}\cS_{2n-1}$, then
$$
\big|\D(E)+1-2\cos^2{\sqrt E\/2}\big|
< 2|\cos^2{\sqrt E\/2}|.
$$
Therefore, the function
$\D(E)+1$ has exactly two zeros in each domain
$\cS_{2n-1},n\in\N$, and has no zeros in the domain
$\cS\sm\cup_{n\in\N}\cS_{2n-1}$.

Thus, the function
$\D^2(E)-1$ has exactly two  zeros in each domain
$\cS_{n},n\in\N$, and has no zeros in the domain
$E\in\cS\sm\cup_{n\in\N}\cS_{n}$.
Therefore, there are exactly two
eigenvalues $E_n^\pm$ of the 2-periodic problem \er{2oequd1}, \er{fbc}
in each domain $\cS_n,n\in\N$, and there are no eigenvalues
 in the domain $\cS\sm\cup_{n\in\N}\cS_{n}$.

Due to \er{PiiimD},
$\D$ is analytic on the half-plane $\Pi_{3\/4}$ and, if $\p$ is real,
then $\D$ is real on the half-line
$[{3\/4},+\iy)$.
The whole spectrum of the 2-periodic problem in $\Pi_2$ is real, then all zeros of the
function $\D^2-1$ in $\Pi_2$ are real.

Moreover,
the spectra of the quasi-periodic problems
\er{2oequd1}, \er{qpbc}
in the half-plane $\Pi_2$ are real, then the standard Krein's arguments
give: if $\D(\l)\in(-1,1)$, then $\D'(\l)\ne 0$,
if $\D(\l)=\pm 1$ and $\D'(\l)=0$, then $\D(\l)\D''(\l)<0$.

These results show that the Lyapunov function
on the half-line $[2,+\iy)$ oscillates in the similar way as in the case
of the Schrodinger operator with a potential, which does not depend on energy.
This yields the statement and the relations \er{Ejpm}.

ii) Similarly, the function $\vp(1,\cdot)$ is analytic in $\cS$
and real on the half-line
$\cS\cup\R$.
If $E\in\cS$, then we have the estimate
$$
\big|\vp(1,E)-{\sin \sqrt E\/\sqrt E}\big|
\le {\| V(\cdot,E)\|\/\sqrt E}e^{\| V(\cdot,E)\|},
$$
see \cite[Lm~3.2~ii)]{BK21}.
Let $E\in\cS\sm\cup_{n\in\N}\cS_{n}$. The definition \er{defOmegan}
gives $|\sqrt E-\pi n|\ge{\sqrt 3\/2}$ for all $n\in\N$. Then
$
|e^{|\Im\sqrt E|}|<C|\sin\sqrt E|,
$
for some $C>0$, see \cite{PT87}.
The estimate \er{estcVr} gives
$$
\Big|\vp(1,E)-{\sin \sqrt E\/\sqrt E}\Big|
\le C\|\p\|_1e^{C\|\p\|_1}e^{|\Im\sqrt E|}
<  \Big|{\sin\sqrt E\/\sqrt E}\Big|.
$$
Then, by Rouche's theorem, the functions $\vp(1,E)$ and
$\sin\sqrt E$ has the same number of zeros in each domain
$\cS_n,n\in\N$, and in the domain $\cS\sm\cup_{n\in\N}\cS_{n}$.
Since the function $\sin\sqrt E$ has exactly one simple zero
at each point $\pi n,n\in\Z$, and has no other zeros, the function
$\vp(1,E)$ has exactly one simple zero in each domain
$\cS_{n},n\in\N$, and has no the zeros in the domain
$\cS\sm\cup_{n\in\N}\cS_{n}$.
If $\p$ is real, then these zeros are real and satisfy
the estimates \er{gmj}.

We have the identities
\[
\lb{D^2}
\begin{aligned}
\D^2(E)=\Big({\vt(1,E)+\vp'(1,E)\/2}\Big)^2
=\Big({\vt(1,E)-\vp'(1,E)\/2}\Big)^2+\vt(1,E)\vp'(1,E)
\\
=\Big({\vt(1,E)-\vp'(1,E)\/2}\Big)^2+\vt'(1,E)\vp(1,E)+1
\end{aligned}
\]
for all $E\in\cS$. Let $E=\gm_n,n\in\N$. Then $\vp(1,E)=0$
and we obtain $\D^2(E)\ge 1$. The relations \er{Ejpm} give
the second relations in \er{gmj}.
~\BBox

\subsection{Results for transpose operator}
Let $\l\in\mD_1$, where
$$
\mD_1=\C_+\sm\cup_{n\in\{0\}\cup\N}\ol{\cD_n}=-\mD_3.
$$
Then Eq.~\er{1btr} has a simple multiplier
$\wt\t_1(\l,\p)=\t_3^{-1}(-\l,\p_*)$
and the corresponding Floquet solution
\[
\lb{symfs3}
\tilde\gf_1(x,\l,\p)=\gf_3(x,-\l,\p_*).
\]
This solution satisfies
$\tilde\gf_1(x,\l)\ne 0$ for all $(x,\l)\in\R\ts\mD_1$,
see \cite{BK24xxx}. Thus, we may use this solution, as $\eta$,
in the substitution \er{deff1} for Eq.~\er{1btr}.

In fact, let $\tilde y$ be a solution of Eq.~\er{1btr} and let
$$
\tilde f=\tilde\gf_1^{3\/2}\Big({\tilde y\/\tilde\gf_1}\Big)'.
$$
Then $(\tilde y^{[2]})'\tilde\gf_1-\tilde y(\tilde\gf_1^{[2]})'
=-p\tilde\gf_1^{1\/2} f$
and we obtain
\[
\lb{Schrforconj}
-\tilde f''+\tilde V\tilde f=E\tilde f,\qq E={3\/4}\l^{2\/3}\in\cS,
\]
where
\[
\lb{deftV}
\tilde V=E-p-{\tilde\gf_1^{[2]}\/\tilde\gf_1}
+{1\/4}\Big({\tilde\gf_1'\/\tilde\gf_1}\Big)^2
-{1\/2}\lt({\tilde\gf_1^{[2]}-p\tilde\gf_1\/\tilde\gf_1}-\Big({\tilde\gf_1'\/\tilde\gf_1}\Big)^2\rt)
=E-{p\/2}-{3\/2}{\tilde\gf_1^{[2]}\/\tilde\gf_1}
+{3\/4}\Big({\tilde\gf_1'\/\tilde\gf_1}\Big)^2.
\]
The identity \er{symfs3} shows that
\[
\lb{symptVwtV}
\tilde V(E,\p_*)=V(E,\p),\qq E\in\cS.
\]

Theorem~\ref{LmtpevschDir} gives the following results.

\begin{corollary}
\lb{Cortpev}

Let $\p\in\cB_\C(\ve_0)$.

i) Let $\l\in\mD_1$ be an eigenvalue of the 3-point Dirichlet problem
for Eq.~\er{1b} and let $y$ be the corresponding eigenfunction.
Then
$E={3\/4}(-\l)^{2\/3}\in\cS$ is an eigenvalue of the Dirichlet problem for
Eq.~\er{Schrforconj} and $\tilde f={y\/\sqrt{\tilde\gf_1}}$
is the corresponding eigenfunction.

ii) Let $E\in\cS$ be an eigenvalue of the Dirichlet problem for
Eq.~\er{Schrforconj} and let $\tilde f$ be the corresponding eigenfunction.
Then $\l=-({4\/3}E)^{3\/2}\in\mD_1$ is  the eigenvalue of the
three-point Dirichlet problem for Eq.~\er{1b}
 and $y=\tilde f\sqrt{\tilde\gf_1}$ is
the corresponding eigenfunction.
\end{corollary}

\subsection{Transformation of ramifications and 3-point eigenvalues}

The following lemma establishes some relations between the set
of the ramifications and three-point eigenvalues for the third order operators
and the set of the 2-periodic and the three-point eigenvalues
for the Schr\"odinger operators.

\begin{lemma}
\lb{Lmev3oand20}

Let $\p\in\cB_\C(\ve_0)$. Then

i) Each number ${3\/4}(r_n^\pm)^{2\/3},n\ge 1$,
is an eigenvalue of the 2-periodic problem for Eq.~\er{2oequd1}.
Each number $({4\/3}E_n^\pm)^{3\/2},n\ge 1$, is a ramification
for Eq.~\er{1b}.

ii) Each number ${3\/4}(-\m_{-n})^{2\/3}\in\cS,n\ge 1 $, is
an eigenvalue of the Dirichlet problem for Eq.~\er{Schrforconj}.
The corresponding eigenfunction has the form $y_{-n}\tilde\gf_1^{-{1\/2}}$,
where $y_{-n}$ is the eigenfunction of the 3-point Dirichlet problem
for Eq.~\er{1b} corresponding to the eigenvalue $\m_{-n}$,
$\tilde\gf_1$ is the Floquet solution of
Eq.~\er{1btr} corresponding to the multiplier $\wt\t_1$.

Conversely, each number $-({4\/3}\gm_n)^{3\/2},n\ge 1$, is an eigenvalue of the
3-point Dirichlet problem for Eq.~\er{2oequd1}
 and $\tilde y_{-n}=\tilde f_n\sqrt{\tilde\gf_1}$ is
the corresponding eigenfunction, where  $\tilde f_n$ is the eigenfunction
of the Dirichlet problem for Eq.~\er{Schrforconj}.

\end{lemma}

\no {\bf Proof.} i) If $\p\in\cB_\C(\ve_0)$ and $\l\in\mD_3$, then
the multiplier $\t_3$ is simple, the
corresponding Floquet solution $\gf_3$ does not vanish for all $x\in\R$,
and $\gf_3^{[2]}(0)\ne 0$, see \cite{BK24xxx}. Then
Theorem~\ref{Lmtpevsch}~i) gives the statement.

ii) Corollary~\ref{Cortpev} gives the statement.~\BBox

\medskip

\no {\bf Proof of Theorem~\ref{ThMcKtr}}.
i) The results are proved in Lemma~\ref{Lmalt2o}~i).

 ii) Lemma~\ref{Lmev3oand20}~i) and the statement i) give the results.~\BBox

 \medskip

\no {\bf Proof of Theorem~\ref{ThMcKtrDir}~i)-ii)}.
i)  The results are proved in Lemma~\ref{Lmalt2o}~ii).

ii) Let $n\in\N$. Lemma~\ref{Lmev3oand20} gives
\er{relevdir-}.
The identities \er{symev} gives
$-\m_{-n}(\p)=\m_n(\p_*^-)$.
The identity \er{epsstar} implies
$$
{3\/4}(\m_n(\p_*^-))^{2\/3}=\gm_n(\p_*),
$$
which yields \er{relr3p2oDir}.

Lemmas~\ref{Lmrealsp}~ii) and \ref{Lmev3oand20}~i) yield that
all $\mu_n$ and $r_n^\pm,n\in\N$, are positive.
Let $n\in\N$. The relations \er{gmj} imply
$$
\gm_n(\p^-)\in[E_n^-(\p^-),E_n^+(\p^-)].
$$
The identities \er{relr3p2o}
and \er{simram} yield
\er{altmurn}.
\BBox

\subsection{Norming constants for three-point problem}
Let $n\in\N$.
Norming constants for the Dirichlet problem for the second order
operator are introduced by
$$
\gh_{sn}=2\pi n\log\big((-1)^n\vp'(1,\gm_n)\big),
$$
see \er{ncschr}.
In the following theorem we express $\gh_{sn}$ in terms of the eigenfunctions
of the 3-point problem.

\begin{proposition}
The function $\vp$ and the norming constants $\gh_{sn}$  satisfy
\[
\lb{vp'hill}
\vp'(1,\gm_n)
={\wt y'(1,\wt\m_n)\/\wt y'(0,\wt\m_n)}\t_3^{-{1\/2}}(\wt\m_n),
\]
\[
\lb{ncwrong}
\gh_{sn}
=2\pi n\log\Big((-1)^n{\wt y'(1,\wt\m_n)\/\wt y'(0,\wt\m_n)}
\t_3^{-{1\/2}}(\wt\m_n)\Big),\qq n\in\N.
\]
where $\wt y(t,\wt\m_n)$ is the eigenfunction of the
operator $\wt\cL_{dir}$, corresponding
to the eigenvalue $\wt\m_n$, $\wt y'(0,\wt\m_n)\ne 0$.
\end{proposition}

\no {\bf Proof.} McKean's transformation gives
$$
\vp(t,\gm_n)={C\wt y(t,\wt\m_n)\/\sqrt{\gf_3(t,\wt\m_n)}},
$$
for some $C>0$.
This yields
$$
\vp'(t,\gm_n)=C\Big({\wt y'(t,\wt\m_n)\/\sqrt{\gf_3(t,\wt\m_n)}}
-{\wt y(t,\wt\m_n)\gf_3'(t,\wt\m_n)\/\gf_3^{3\/2}(t,\wt\m_n)}\Big).
$$
Then, using $\wt y(0,\wt\m_n)=\wt y(1,\wt\m_n)=\wt y(2,\wt\m_n)=0$, we obtain
$$
\vp'(0,\gm_n)={C\wt y'(0,\wt\m_n)\/\sqrt{\gf_3(0,\wt\m_n)}},\qq
\vp'(1,\gm_n)={C\wt y'(1,\wt\m_n)\/\sqrt{\gf_3(1,\wt\m_n)}}.
$$
The identity $\vp'(0,\gm_n)=1$ gives
$$
C={\sqrt{\gf_3(0,\wt\m_n)}\/\wt y'(0,\wt\m_n)},
$$
which yields
$$
\vp'(1,\gm_n)
={\wt y'(1,\wt\m_n)\sqrt{\gf_3(0,\wt\m_n)}\/\wt y'(0,\wt\m_n)
\sqrt{\gf_3(1,\wt\m_n)}}.
$$
This gives \er{vp'hill}, which implies \er{ncwrong}.~\BBox

\medskip

\no {\bf Proof of Theorem~\ref{ThMcKtrDir}~iii).}
The definition \er{defnf} and the identity \er{ncwrong}
imply \er{nc2o}.~\BBox

\section{Riemann surfaces}
\setcounter{equation}{0}

\subsection{Riemann surfaces for the case $\p=\const$}

Assume that $q=0$ and $p=p_0=\const$. In this case
the multipliers have the form
$\t_j=e^{k_j},j=1,2,3$, where $k_j$ are three zeros of the polynomial $k^3+2p_0k-\l$.
The multiplier Riemann surface $\cR$ coincides with the surface defined by the algebraic
equation $k^3+2p_0k-\l=0$. There are exactly two branch points
$r_0^\pm=\pm{4\/3}\sqrt{2\/3}(-p_0)^{3\/2}$. The other ramifications
$$
r_n^+=r_n^-={4\/3\sqrt3}\big(2(\pi n)^2-p_0\big)\big((\pi n)^2-2p_0\big)^{1\/2},
\qq r_{-n}^\pm=-r_n^{\pm},\qq n\in\N,
$$
have multiplicity two and the corresponding branch points are degenerated.

Assume $p_0<0$ for definiteness, then $r_0^+=-r_0^->0$. In order to construct the surface $\cR$
we take three replicas $\cR_1,\cR_2$ and $\cR_3$
of the slitted complex plane:
$$
\cR_1=\C\sm[r_0^+,+\iy),\qq
\cR_2=\C\sm\big((-\iy,r_0^-]\cup_{n\in\Z}[r_0^+,+\iy)\big),\qq
\cR_3=\C\sm(-\iy,r_0^-].
$$
We attach the upper (lower) edge of the slit $[r_0^+,+\iy)$ on the sheet $\cR_1$
to the lower (upper) edge of the same slit on the sheet $\cR_2$.
Similarly, the sheets $\cR_2$ and $\cR_3$ of the surface $\cR$
are attached along the slits $(-\iy,r_0^-]$, see  Fig.~\ref{FigrsZEL1}~c).

Consider the parametrization of the surface $\cR$
by the analytical mapping $\cZ(\l)=\l^{1\/3},\l\in\cR$, Fig.~\ref{FigrsZEL1}~a).
We have $\cZ(\cR)=\cZ$, where
$$
\cZ=\C\sm\cup_{j\in\{0,2,3,5\}}[0,z_{0,j}],\qq z_{0,j}=(r_0^+)^{1\/3}e^{i{\pi\/3}j},\qq
1^{1\/3}=1.
$$
On the plane $\cZ$ we identify with each other
the corresponding edges of the slits
$[0,z_{0,j}]$, $n\in\N$, $j=0$ with $j=2$
and $j=3$ with $j=5$: we identify the upper edge of the first slit
with the upper edge of the second slit, we identify the lower edge
of the first slit with the lower edge of the second slit.
Moreover, we have
$\cZ(\cR_j^{\pm})=\cZ_j^{\pm},j=1,2,3$, where $\cR_j^{\pm}$
is a part of the sheet $\cR_j$ over $\C_\pm$, and $\cZ_j^{\pm}$
is the sector shown in Fig.~\ref{FigrsZEL1}~a).

The Floquet solutions have the form
$\gf_j=e^{k_j x},j=1,2,3$.
For each fixed $x\in\R$ the functions $\gf_j(x,\cdot)$ constitute three branches of a function
$\gf(x,\cdot)$ analytic on the surface $\cR$.
The definition \er{da1+} gives $V=E-2p_0-{3\/4}k^2$ and it is analytic function
on the surface $\cE$, which is the Riemann surface of the function $k(\l(E)),\l(E)=({4\/3}E)^{3\/2}$.

In order to construct the surface $\cE$, see Fig.~\ref{FigrsZEL1}~b),
we take two replicas $\cE_1$ and $\cE_2$
of the slitted complex plane:
$$
\cE_1=\cE_2=\C\sm\Big((-\iy,0]\bigcup(\cup_{j\in\{0,2\}}
[0,E_0^+]e^{i{2\pi\/3}j})\Big),\qq E_0^+={3\/4}(r_0^+)^{2\/3}=-{p_0\/\sqrt[3]2}.
$$
We attach the upper (lower) edge of the slit $(-\iy,0]$, on the sheet $\cE_1$
to the lower (upper) edge of the same slit on the sheet $\cE_2$.
We identify with each other
the corresponding edges of the slits
$[0,E_0^+]$ on the sheet $\cE_1$
with the slits $[0,E_0^+]e^{i{4\pi\/3}}$ on the sheet $\cE_2$:
we identify the upper edge of the first slit
with the upper edge of the second slit, we identify the lower edge
of the first slit with the lower edge of the second slit.
Similarly, we identify
the corresponding edges of the slits
$[0,E_0^+]e^{i{4\pi\/3}}$ on the sheet $\cE_1$
with the slits $[0,E_0^+]$ on the sheet $\cE_2$.
Moreover, we have
$\cE(\cR_j^{\pm})=\cE_j^{\pm},j=1,2,3$, where $\cE_j^{\pm}$
are the sectors shown in Fig.~\ref{FigrsZEL1}~b).
The points $E_n^+e^{i{2\pi\/3}j},j=0,2,n\in\N,E_n^+={3\/4}(r_n^+)^{2\/3}$,
are degenerated ramifications of the surface $\cE$. The small perturbation divides
these points into pares of simple ramifications.

\begin{figure}
\tiny
\unitlength 0.8mm
\linethickness{0.4pt}
\ifx\plotpoint\undefined\newsavebox{\plotpoint}\fi 

\caption{\footnotesize a) The multipliers $z$-plane, b) $E$-surface $\cE$,
and c) $\l$-surface $\cR$ for the
case $p=\const<0$. Each double ramification $r_n^-=r_n^+$ on the surface $\cR$
(and the corresponding ramifications $z_{n,j}=(r_n^+)^{1\/3}e^{i{\pi\/3}j}$ on the plane $\cZ$
and  $E_ne^{i{\pi\/3}j},E_n={3\/4}(r_n^+)^{2\/3}$, on the surface $\cE$)
divides into two simple ramifications on the surface $\cR$ at small perturbation,
see Fig.~\ref{FigrsZEL}~c)}
\lb{FigrsZEL1}
\end{figure}

Let $p_0\to 0$. Then $r_0^\pm\to 0,r_n^\pm\to(\pi n)^3,n\in\N$,
and in the limit case we obtain the surfaces
for the unperturbed case $p_0=0$.
The multiplier surface $\cR^0$ for the unperturbed case
is the surface of the function $\l^{1\/3}$ and it is shown in Fig.~\ref{FigrsZEL0}~c).
The mapping $z(\l)=\l^{1\/3}$ gives the simple parametrization $\cR^0\to\cZ^0=\C$.
Moreover, we have $\cZ^0(\cR_j^{0,\pm})=\cZ_j^{\pm},j=1,2,3$, where $\cR_j^{0,\pm}$
is a part of the sheet $\cR_j^0$ over $\C_\pm$, and $\cZ_j^{\pm}$
are the sectors shown in Fig.~\ref{FigrsZEL0}~a).
The corresponding Riemann
surface $\cE^0$ of the function $k(\l(E))$ is the surface of the function $E^{1\/2}$,
see Fig.~\ref{FigrsZEL0}~b). Note that in this unperturbed case we have $V=0$.

The function $f_j=\gf_3^{3\/2}({\gf_j\/\gf_3})',j=1,2$,
is the Floquet solution of Eq.~\er{2oequd1} satisfying
$f_j(x+1,E)=t_j(E)f_j(x,E)$,
where $t_j=\tau_3^{1\/2}\tau_j$ is a multiplier for  Eq.~\er{2oequd1}.
Then the Lyapunov function satisfies
$$
\D={1\/2}(t_1+t_2)={1\/2}(\tau_1+\tau_2)\tau_3^{1\/2}={1\/2}(T-\tau_3)\tau_3^{1\/2},
$$
where $T(E)=\Tr M(1,\l(E))$. Since $\Tr M(1,\cdot)$ is an entire function,
the function $\D$ is analytic on the surface $\cE$.

\begin{figure}
\tiny
\unitlength 0.8mm
\linethickness{0.4pt}
\ifx\plotpoint\undefined\newsavebox{\plotpoint}\fi 
\begin{picture}(198.879,135)(0,0)
\put(76.85,112.925){\makebox(0,0)[cc]{$\cZ_1^+$}}
\put(19.6,114.45){\makebox(0,0)[cc]{$\cZ_2^+$}}
\put(21.55,78.05){\makebox(0,0)[cc]{$\cZ_3^-$}}
\put(74.175,78.475){\makebox(0,0)[cc]{$\cZ_2^-$}}
\multiput(129.654,47.665)(.03656462585,-.03373015873){882}{\line(1,0){.03656462585}}
\multiput(129.654,85.415)(.03656462585,-.03373015873){882}{\line(1,0){.03656462585}}
\multiput(129.654,123.165)(.03656462585,-.03373015873){882}{\line(1,0){.03656462585}}
\multiput(166.654,44.415)(.03920220083,-.03370013755){727}{\line(1,0){.03920220083}}
\multiput(94.154,118.915)(.03920220083,-.03370013755){727}{\line(1,0){.03920220083}}
\multiput(109.144,31.331)(-.03355932,.66313559){59}{\line(0,1){.66313559}}
\multiput(94.221,44.079)(.038,-.0337084548){343}{\line(1,0){.038}}
\multiput(109.358,31.045)(.0393776435,-.0336586103){331}{\line(1,0){.0393776435}}
\multiput(166.806,118.499)(.0452013652,-.0337201365){293}{\line(1,0){.0452013652}}
\multiput(181.101,107.147)(.0376604278,-.0337245989){374}{\line(1,0){.0376604278}}
\multiput(94.432,81.5)(.0394553846,-.0336369231){325}{\line(1,0){.0394553846}}
\multiput(108.727,69.096)(.0391547278,-.033730659){349}{\line(1,0){.0391547278}}
\put(198.879,70.358){\line(0,1){0}}
\put(145.74,69.789){\line(0,-1){.9588}}
\put(145.844,67.871){\line(0,-1){.9588}}
\put(145.948,65.954){\line(0,-1){.9588}}
\put(146.052,64.036){\line(0,-1){.9588}}
\put(146.156,62.118){\line(0,-1){.9588}}
\put(146.26,60.201){\line(0,-1){.9588}}
\put(146.364,58.283){\line(0,-1){.9588}}
\put(146.468,56.366){\line(0,-1){.9588}}
\put(146.572,54.448){\line(0,-1){.9588}}
\put(146.676,52.53){\line(0,-1){.9588}}
\multiput(146.702,50.386)(.03302222,-.43935556){45}{\line(0,-1){.43935556}}
\put(146.632,48.235){\line(0,-1){.9941}}
\put(146.502,46.246){\line(0,-1){.9941}}
\put(146.371,44.258){\line(0,-1){.9941}}
\put(146.241,42.27){\line(0,-1){.9941}}
\put(146.111,40.282){\line(0,-1){.9941}}
\put(145.981,38.293){\line(0,-1){.9941}}
\put(145.851,36.305){\line(0,-1){.9941}}
\put(145.721,34.317){\line(0,-1){.9941}}
\multiput(146.053,89.353)(-.0323478,-.8143478){23}{\line(0,-1){.8143478}}
\multiput(179.764,108.374)(.0296,-.2676){5}{\line(0,-1){.2676}}
\put(179.694,106.966){\line(0,-1){.9719}}
\put(179.762,105.022){\line(0,-1){.9719}}
\put(179.831,103.078){\line(0,-1){.9719}}
\put(179.9,101.134){\line(0,-1){.9719}}
\put(179.968,99.19){\line(0,-1){.9719}}
\put(180.037,97.246){\line(0,-1){.9719}}
\put(180.105,95.303){\line(0,-1){.9719}}
\put(181.031,107.115){\line(0,-1){.972}}
\put(180.962,105.171){\line(0,-1){.972}}
\put(180.893,103.227){\line(0,-1){.972}}
\put(180.825,101.283){\line(0,-1){.972}}
\put(180.756,99.339){\line(0,-1){.972}}
\put(180.688,97.395){\line(0,-1){.972}}
\put(180.619,95.451){\line(0,-1){.972}}
\put(180.655,94.401){\line(0,-1){6.987}}
\put(180.437,87.047){\line(0,-1){.9531}}
\put(180.367,85.14){\line(0,-1){.9531}}
\put(180.297,83.234){\line(0,-1){.9531}}
\put(180.227,81.328){\line(0,-1){.9531}}
\put(180.157,79.422){\line(0,-1){.9531}}
\put(180.087,77.516){\line(0,-1){.9531}}
\put(180.017,75.609){\line(0,-1){.9531}}
\put(179.947,73.703){\line(0,-1){.9531}}
\put(179.877,71.797){\line(0,-1){.9531}}
\put(103.125,115.8){\makebox(0,0)[cc]{$\cR_1^0$}}
\put(103.125,78.785){\makebox(0,0)[cc]{$\cR_2^0$}}
\put(103.125,41.474){\makebox(0,0)[cc]{$\cR_3^0$}}
\put(147.043,111.361){\makebox(0,0)[cc]{$0$}}
\put(161.009,111.769){\makebox(0,0)[cc]{$\pi^3$}}
\put(151.107,30.426){\makebox(0,0)[cc]{$0$}}
\put(132.573,29.173){\makebox(0,0)[cc]{$-\pi^3$}}
\put(72.125,97.25){\line(1,0){.25}}
\multiput(146.375,94.25)(-.03125,-.5){8}{\line(0,-1){.5}}
\put(147.555,60.93){\line(0,1){.9444}}
\put(147.61,62.819){\line(0,1){.9444}}
\put(147.666,64.707){\line(0,1){.9444}}
\put(147.721,66.596){\line(0,1){.9444}}
\put(147.777,68.485){\line(0,1){.9444}}
\put(11.125,96.75){\line(1,0){75.5}}
\multiput(31.375,70.75)(.0337398374,.04695121951){1230}{\line(0,1){.04695121951}}
\multiput(23.875,132.75)(.03371833085,-.0467585693){1342}{\line(0,-1){.0467585693}}
\put(94.125,118.75){\line(1,0){72.5}}
\put(106.875,107.75){\line(1,0){73}}
\put(122.375,94.25){\line(1,0){73}}
\put(108.125,69.75){\line(1,0){73}}
\put(122.625,57){\line(1,0){72.75}}
\multiput(94.375,44.25)(1.6875,-.03125){8}{\line(1,0){1.6875}}
\put(107.625,32.25){\line(1,0){73}}
\put(122.375,19.75){\line(1,0){72.5}}
\put(8.625,54.5){\line(1,0){68.25}}
\put(16.625,44.5){\line(1,0){68.25}}
\put(16.625,15){\line(1,0){68.25}}
\put(24.625,34.5){\line(1,0){68.25}}
\put(24.625,5){\line(1,0){68.25}}
\multiput(76.625,54.5)(.0336842105,-.0426315789){475}{\line(0,-1){.0426315789}}
\multiput(76.625,25)(.0336842105,-.0426315789){475}{\line(0,-1){.0426315789}}
\multiput(50.125,44.5)(-.0867003367,.0336700337){297}{\line(-1,0){.0867003367}}
\multiput(50.125,15)(-.0867003367,.0336700337){297}{\line(-1,0){.0867003367}}
\multiput(50.125,44)(-.0409090909,-.0336363636){275}{\line(-1,0){.0409090909}}
\multiput(50.125,14.5)(-.0409090909,-.0336363636){275}{\line(-1,0){.0409090909}}
\put(15.625,45.5){\line(1,0){34}}
\put(17.375,43.25){\line(1,0){33.5}}
\put(17.375,13.75){\line(1,0){33.5}}
\multiput(49.125,45.75)(.03333333,-.03333333){75}{\line(0,-1){.03333333}}
\multiput(15.625,45.5)(.03365385,-.62019231){52}{\line(0,-1){.62019231}}
\multiput(17.625,43)(-.03333333,-.45416667){60}{\line(0,-1){.45416667}}
\put(49.555,45.68){\line(0,-1){.9792}}
\put(49.638,43.721){\line(0,-1){.9792}}
\put(49.721,41.763){\line(0,-1){.9792}}
\put(49.805,39.805){\line(0,-1){.9792}}
\put(49.888,37.846){\line(0,-1){.9792}}
\put(49.971,35.888){\line(0,-1){.9792}}
\put(8.625,24.75){\line(1,0){7.75}}
\put(16.805,24.68){\line(1,0){.9786}}
\put(18.762,24.694){\line(1,0){.9786}}
\put(20.719,24.708){\line(1,0){.9786}}
\put(22.676,24.723){\line(1,0){.9786}}
\put(24.633,24.737){\line(1,0){.9786}}
\put(26.59,24.751){\line(1,0){.9786}}
\put(28.548,24.765){\line(1,0){.9786}}
\put(30.505,24.78){\line(1,0){.9786}}
\put(32.462,24.794){\line(1,0){.9786}}
\put(34.419,24.808){\line(1,0){.9786}}
\put(36.376,24.823){\line(1,0){.9786}}
\put(38.333,24.837){\line(1,0){.9786}}
\put(40.29,24.851){\line(1,0){.9786}}
\put(42.248,24.865){\line(1,0){.9786}}
\put(44.205,24.88){\line(1,0){.9786}}
\put(46.162,24.894){\line(1,0){.9786}}
\put(48.119,24.908){\line(1,0){.9786}}
\put(50.076,24.923){\line(1,0){.9786}}
\put(51.125,25.25){\line(1,0){25.75}}
\multiput(8.375,54.5)(.03372093,-.043023256){215}{\line(0,-1){.043023256}}
\multiput(17.875,42.5)(.033678756,-.044041451){193}{\line(0,-1){.044041451}}
\multiput(8.625,24.75)(.033678756,-.044041451){193}{\line(0,-1){.044041451}}
\multiput(17.125,13.75)(.03372093,-.040697674){215}{\line(0,-1){.040697674}}
\put(60.625,50.75){\makebox(0,0)[cc]{$\cE_1^{0,+}$}}
\put(16.375,51){\makebox(0,0)[cc]{$\cE_{1,r}^{0,-}$}}
\put(26.375,7.75){\makebox(0,0)[cc]{$\cE_{1,l}^{0,-}$}}
\put(86.125,8){\makebox(0,0)[cc]{$\cE_2^{0,+}$}}
\put(69.125,22.75){\makebox(0,0)[cc]{$\cE_3^{0,-}$}}
\put(21.125,21){\makebox(0,0)[cc]{$\cE_{3,l}^{0,+}$}}
\put(27.875,38.25){\makebox(0,0)[cc]{$\cE_{3,r}^{0,+}$}}
\put(62.375,38){\makebox(0,0)[cc]{$\cE_2^{0,-}$}}
\put(53.125,62.75){\makebox(0,0)[cc]{a)}}
\put(8.625,5.75){\makebox(0,0)[cc]{b)}}
\put(146.375,9){\makebox(0,0)[cc]{c)}}
\put(145.055,108.43){\line(0,-1){.95}}
\put(145.155,106.53){\line(0,-1){.95}}
\put(145.255,104.63){\line(0,-1){.95}}
\put(145.355,102.73){\line(0,-1){.95}}
\put(145.455,100.83){\line(0,-1){.95}}
\put(145.555,98.93){\line(0,-1){.95}}
\put(145.655,97.03){\line(0,-1){.95}}
\put(145.755,95.13){\line(0,-1){.95}}
\put(147.555,106.18){\line(0,-1){.9423}}
\put(147.401,104.295){\line(0,-1){.9423}}
\put(147.247,102.41){\line(0,-1){.9423}}
\put(147.093,100.526){\line(0,-1){.9423}}
\put(146.939,98.641){\line(0,-1){.9423}}
\put(146.785,96.757){\line(0,-1){.9423}}
\put(146.632,94.872){\line(0,-1){.9423}}
\multiput(147.375,56.75)(-.0326087,-.3043478){23}{\line(0,-1){.3043478}}
\put(147.555,60.93){\line(0,-1){.85}}
\put(147.455,59.23){\line(0,-1){.85}}
\put(147.355,57.53){\line(0,-1){.85}}
\put(147.875,44.25){\line(1,0){19}}
\put(15.305,16.18){\line(1,0){.9714}}
\put(17.248,16.18){\line(1,0){.9714}}
\put(19.19,16.18){\line(1,0){.9714}}
\put(21.133,16.18){\line(1,0){.9714}}
\put(23.076,16.18){\line(1,0){.9714}}
\put(25.019,16.18){\line(1,0){.9714}}
\put(26.962,16.18){\line(1,0){.9714}}
\put(28.905,16.18){\line(1,0){.9714}}
\put(30.848,16.18){\line(1,0){.9714}}
\put(32.79,16.18){\line(1,0){.9714}}
\put(34.733,16.18){\line(1,0){.9714}}
\put(36.676,16.18){\line(1,0){.9714}}
\put(38.619,16.18){\line(1,0){.9714}}
\put(40.562,16.18){\line(1,0){.9714}}
\put(42.505,16.18){\line(1,0){.9714}}
\put(44.448,16.18){\line(1,0){.9714}}
\put(46.39,16.18){\line(1,0){.9714}}
\put(48.333,16.18){\line(1,0){.9714}}
\put(50.555,28.68){\line(0,-1){.9615}}
\put(50.401,26.757){\line(0,-1){.9615}}
\put(50.247,24.834){\line(0,-1){.9615}}
\put(50.093,22.91){\line(0,-1){.9615}}
\put(49.939,20.987){\line(0,-1){.9615}}
\put(49.785,19.064){\line(0,-1){.9615}}
\put(49.632,17.141){\line(0,-1){.9615}}
\multiput(49.055,16.43)(.0323529,-.0323529){17}{\line(0,-1){.0323529}}
\multiput(50.155,15.33)(.0323529,-.0323529){17}{\line(1,0){.0323529}}
\multiput(51.255,14.23)(.0323529,-.0323529){17}{\line(0,-1){.0323529}}
\multiput(51.125,34.5)(-.0333333,-.3833333){15}{\line(0,-1){.3833333}}
\multiput(50.625,28.25)(.0333333,-.4833333){30}{\line(0,-1){.4833333}}
\put(50.125,65.75){\line(0,1){69.25}}
\put(58.375,130.25){\makebox(0,0)[cc]{$\cZ_{1,r}^-$}}
\put(37.625,130.75){\makebox(0,0)[cc]{$\cZ_{1,l}^-$}}
\put(37.375,69){\makebox(0,0)[cc]{$\cZ_{3,l}^+$}}
\put(56.125,69){\makebox(0,0)[cc]{$\cZ_{3,r}^+$}}
\put(148.125,106.25){\line(1,0){33.75}}
\put(147.125,68.75){\line(1,0){34.5}}
\put(145.555,70.93){\line(1,0){.9786}}
\put(147.512,70.915){\line(1,0){.9786}}
\put(149.469,70.901){\line(1,0){.9786}}
\put(151.426,70.887){\line(1,0){.9786}}
\put(153.383,70.873){\line(1,0){.9786}}
\put(155.34,70.858){\line(1,0){.9786}}
\put(157.298,70.844){\line(1,0){.9786}}
\put(159.255,70.83){\line(1,0){.9786}}
\put(161.212,70.815){\line(1,0){.9786}}
\put(163.169,70.801){\line(1,0){.9786}}
\put(165.126,70.787){\line(1,0){.9786}}
\put(167.083,70.773){\line(1,0){.9786}}
\put(169.04,70.758){\line(1,0){.9786}}
\put(170.998,70.744){\line(1,0){.9786}}
\put(172.955,70.73){\line(1,0){.9786}}
\put(174.912,70.715){\line(1,0){.9786}}
\put(176.869,70.701){\line(1,0){.9786}}
\put(178.826,70.687){\line(1,0){.9786}}
\put(108.75,68.75){\line(1,0){38.75}}
\put(107.125,71){\line(1,0){39}}
\put(93.625,81.25){\line(1,0){52}}
\put(145.555,82.18){\line(1,0){.9659}}
\put(147.487,82.157){\line(1,0){.9659}}
\put(149.418,82.134){\line(1,0){.9659}}
\put(151.35,82.112){\line(1,0){.9659}}
\put(153.282,82.089){\line(1,0){.9659}}
\put(155.214,82.066){\line(1,0){.9659}}
\put(157.146,82.043){\line(1,0){.9659}}
\put(159.077,82.021){\line(1,0){.9659}}
\put(161.009,81.998){\line(1,0){.9659}}
\put(162.941,81.975){\line(1,0){.9659}}
\put(164.873,81.952){\line(1,0){.9659}}
\multiput(166.305,81.68)(.0416667,-.033179){18}{\line(1,0){.0416667}}
\multiput(167.805,80.485)(.0416667,-.033179){18}{\line(1,0){.0416667}}
\multiput(169.305,79.291)(.0416667,-.033179){18}{\line(1,0){.0416667}}
\multiput(170.805,78.096)(.0416667,-.033179){18}{\line(1,0){.0416667}}
\multiput(172.305,76.902)(.0416667,-.033179){18}{\line(1,0){.0416667}}
\multiput(173.805,75.707)(.0416667,-.033179){18}{\line(1,0){.0416667}}
\multiput(175.305,74.513)(.0416667,-.033179){18}{\line(1,0){.0416667}}
\multiput(176.805,73.319)(.0416667,-.033179){18}{\line(1,0){.0416667}}
\multiput(178.305,72.124)(.0416667,-.033179){18}{\line(1,0){.0416667}}
\multiput(181.625,68.5)(.0401146132,-.0336676218){349}{\line(1,0){.0401146132}}
\put(107.805,43.805){\line(1,0){.9813}}
\put(109.767,43.805){\line(1,0){.9813}}
\put(111.73,43.805){\line(1,0){.9813}}
\put(113.692,43.805){\line(1,0){.9813}}
\put(115.655,43.805){\line(1,0){.9812}}
\put(117.617,43.805){\line(1,0){.9813}}
\put(119.58,43.805){\line(1,0){.9813}}
\put(121.542,43.805){\line(1,0){.9812}}
\put(123.505,43.805){\line(1,0){.9813}}
\put(125.467,43.805){\line(1,0){.9813}}
\put(127.43,43.805){\line(1,0){.9812}}
\put(129.392,43.805){\line(1,0){.9812}}
\put(131.355,43.805){\line(1,0){.9813}}
\put(133.317,43.805){\line(1,0){.9813}}
\put(135.28,43.805){\line(1,0){.9812}}
\put(137.242,43.805){\line(1,0){.9812}}
\put(139.205,43.805){\line(1,0){.9812}}
\put(141.167,43.805){\line(1,0){.9813}}
\put(143.13,43.805){\line(1,0){.9812}}
\put(145.092,43.805){\line(1,0){.9813}}
\put(109.25,30.875){\line(1,0){39}}
\put(106.18,33.18){\line(1,0){.975}}
\put(108.13,33.18){\line(1,0){.975}}
\put(110.08,33.18){\line(1,0){.975}}
\put(112.03,33.18){\line(1,0){.975}}
\put(113.98,33.18){\line(1,0){.975}}
\put(115.93,33.18){\line(1,0){.975}}
\put(117.88,33.18){\line(1,0){.975}}
\put(119.83,33.18){\line(1,0){.975}}
\put(121.78,33.18){\line(1,0){.975}}
\put(123.73,33.18){\line(1,0){.975}}
\put(125.68,33.18){\line(1,0){.975}}
\put(127.63,33.18){\line(1,0){.975}}
\put(129.58,33.18){\line(1,0){.975}}
\put(131.53,33.18){\line(1,0){.975}}
\put(133.48,33.18){\line(1,0){.975}}
\put(135.43,33.18){\line(1,0){.975}}
\put(137.38,33.18){\line(1,0){.975}}
\put(139.33,33.18){\line(1,0){.975}}
\put(141.28,33.18){\line(1,0){.975}}
\put(143.23,33.18){\line(1,0){.975}}
\put(145.125,108.875){\line(1,0){34.375}}
\multiput(180.25,94.125)(.03308824,-.73897059){34}{\line(0,-1){.73897059}}
\multiput(145.75,94.25)(.03333333,-.42916667){60}{\line(0,-1){.42916667}}
\multiput(109,68.875)(-.03333333,-.48){75}{\line(0,-1){.48}}
\put(49.93,34.43){\line(0,-1){.9643}}
\put(50.144,32.501){\line(0,-1){.9643}}
\put(50.358,30.573){\line(0,-1){.9643}}
\put(50.573,28.644){\line(0,-1){.9643}}
\put(51.68,43.18){\line(0,-1){.9583}}
\put(51.513,41.263){\line(0,-1){.9583}}
\put(51.346,39.346){\line(0,-1){.9583}}
\put(51.18,37.43){\line(0,-1){.9583}}
\put(51.013,35.513){\line(0,-1){.9583}}
\multiput(162,108.875)(.03333333,-.03333333){75}{\line(0,-1){.03333333}}
\multiput(130.75,30.625)(-.03333333,.03333333){75}{\line(0,1){.03333333}}
\multiput(162,71.125)(.03333333,-.03333333){75}{\line(0,-1){.03333333}}
\multiput(130.75,68.5)(-.03333333,.03333333){75}{\line(0,1){.03333333}}
\put(71.5,97.75){\line(0,-1){2}}
\put(28.5,95.5){\line(0,1){2}}
\multiput(36.75,113.625)(.0458333,.0333333){30}{\line(1,0){.0458333}}
\multiput(63.25,79.625)(-.0458333,-.0333333){30}{\line(-1,0){.0458333}}
\put(70.75,99.75){\makebox(0,0)[cc]{$\pi$}}
\put(32,113.5){\makebox(0,0)[cc]{$\pi e^{i{2\pi\/3}}$}}
\put(28.125,93.5){\makebox(0,0)[cc]{$-\pi$}}
\put(56.7,77.75){\makebox(0,0)[cc]{$-\pi e^{i{2\pi\/3}}$}}
\multiput(67.375,45.375)(.03289474,-.03947368){38}{\line(0,-1){.03947368}}
\multiput(67.375,15.875)(.03289474,-.03947368){38}{\line(0,-1){.03947368}}
\put(41.375,37.625){\line(1,0){2.5}}
\put(41.375,8.125){\line(1,0){2.5}}
\put(68.75,12){\makebox(0,0)[cc]{${3\/4}\pi^2$}}
\put(52.25,8){\makebox(0,0)[cc]{${3\/4}\pi^2e^{i{4\pi\/3}}$}}
\end{picture}
\caption{\footnotesize a) The multipliers $z$-plane, b) $E$-surface $\cE$,
and c) $\l$-surface $\cR$ for the
case $\p=0$. Each double ramification $r_n^0=(\pi n)^3$ on the surface $\cR^0$
divides into two simple ramifications on the surface $\cR$ at small perturbation,
see Fig.~\ref{FigrsZEL}~c)}
\lb{FigrsZEL0}
\end{figure}
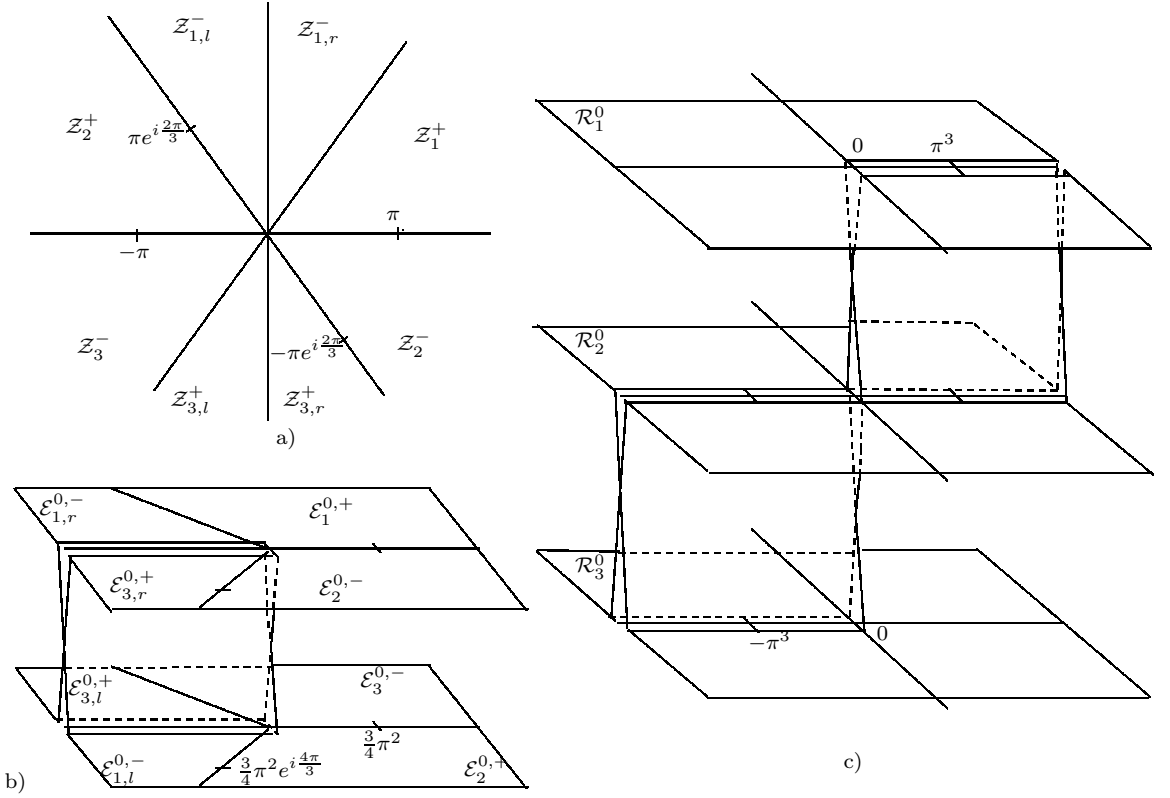

\subsection{Riemann surfaces for the case of small coefficients}
Let $\p\in\cH\os\cH$.
Recall that for each $x\in\R$
the functions $\gf_{1}(x,\cdot),\gf_{2}(x,\cdot),\gf_{3}(x,\cdot)$
constitute three branches of a function $\gf(x,\cdot)$
meromorphic on the multiplier Riemann surface $\cR$ and the poles are not
depending on $x$.

If, in addition,  $\|\p\|\le\ve_0$, then
for any $\l\in\mD_3$ we may take $g=\gf_3$ in
McKean's transformation \er{deff1}--\er{da1+}.
Then each of the potentials $V(x,\cdot),x\in\R$, is analytic on the domain
$\cS$ given by \er{mDBe}. Moreover, it has a meromorphic extension onto
the Riemann surface $\cE$, which is a perturbation of the surface $\cE^0$.

Describe the surface $\cE$.
Represent the construction of $\cR$ from our paper \cite{BK24xxx}
for the case of small coefficients, see Fig.~\ref{FigrsZEL}~c).
Assume $r_0^-=r_0^+=0$ for simplicity.
We take three replicas $\cR_1,\cR_2$ and $\cR_3$
of the slitted complex plane:
$$
\cR_1=\C\sm\cup_{n\in\N}[r_{n-1}^+,r_n^-],\qq
\cR_2=\C\sm\cup_{n\in\Z}[r_{n-1}^+,r_n^-],\qq
\cR_3=\C\sm\cup_{n\le 0}[r_{n-1}^+,r_n^-].
$$
We attach the upper (lower) edge of the slit $[r_{n-1}^+,r_n^-],n\in\N$, on the sheet $\cR_1$
to the lower (upper) edge of the same slit on the sheet $\cR_2$.
Similarly, the sheets $\cR_2$ and $\cR_3$ of the surface $\mR$
are attached along the slits $[r_{n-1}^+,r_n^-],n\le 0$.

Consider the parametrization of the surface $\cR$
by the analytical mapping $\cZ(\l)=\l^{1\/3},\l\in\cR$, Fig.~\ref{FigrsZEL}~a).
We have $\cZ(\cR)=\cZ$, where
$$
\cZ=\C\sm\cup_{n\in\N,j\in\{0,2,3,5\}}[(r_n^-)^{1\/3},(r_n^+)^{1\/3}]e^{i{\pi\/3}j},\qq
1^{1\/3}=1.
$$
On the plane $\cZ$ we identify with each other
the corresponding edges of the slits
$[(r_n^-)^{1\/3},(r_n^+)^{1\/3}]e^{i{\pi\/3}j}$, $n\in\N$, $j=0$ with $j=2$
and $j=3$ with $j=5$: we identify the upper edge of the first slit
with the upper edge of the second slit, we identify the lower edge
of the first slit with the lower edge of the second slit.
Moreover, we have
$\cZ(\cR_j^{\pm})=\cZ_j^{\pm},j=1,2,3$, where $\cR_j^{\pm}$
is a part of the sheet $\cR_j$ over $\C_\pm$, and $\cZ_j^{\pm}$
is the sector shown in Fig.~\ref{FigrsZEL}~a).

The definition \er{da1+} yields that
for each $x\in\R$ the function $V(x,\cdot)$ has a meromorphic
extension from the domain $\cS$ onto a Riemann surface
$\cE$.
Using the identity \er{defEen} we obtain the following form for the surface
$\cE$, see Fig.~\ref{FigrsZEL}~b).
In order to construct the surface $\cE$
we take two replicas $\cE_1$ and $\cE_2$
of the slitted complex plane:
$$
\begin{aligned}
\cE_1=\C\sm\Big((-\iy,0]\bigcup\cup_{n\in\N}
\big([E_n^-(\p^-),E_n^+(\p^-)]\cup [E_n^-(\p_*),E_n^+(\p_*)]e^{i{4\pi\/3}}\big)\Big),\\
\cE_2=\C\sm\Big((-\iy,0]\bigcup\cup_{n\in\N}
\big([E_n^-(\p_*),E_n^+(\p_*)]\cup [E_n^-(\p^-),E_n^+(\p^-)]e^{i{4\pi\/3}}\big)\Big).
\end{aligned}
$$
We attach the upper (lower) edge of the slit $(-\iy,0]$, on the sheet $\cE_1$
to the lower (upper) edge of the same slit on the sheet $\cE_2$.
We identify with each other
the corresponding edges of the slits
$[E_n^-(\p^-),E_n^+(\p^-)]$, $n\in\N$, on the sheet $\cE_1$
with the slits $[E_n^-(\p^-),E_n^+(\p^-)]e^{i{4\pi\/3}}$, $n\in\N$, on the sheet $\cE_2$:
we identify the upper edge of the first slit
with the upper edge of the second slit, we identify the lower edge
of the first slit with the lower edge of the second slit.
Similarly, we identify
the corresponding edges of the slits
$[E_n^-(\p_*),E_n^+(\p_*)]e^{i{4\pi\/3}}$, $n\in\N$, on the sheet $\cE_1$
with the slits $[E_n^-(\p_*),E_n^+(\p_*)]$, $n\in\N$, on the sheet $\cE_2$.
Moreover, we have
$\cE(\cR_j^{\pm})=\cE_j^{\pm},j=1,2,3$, where $\cE_j^{\pm}$
are the sectors shown in Fig.~\ref{FigrsZEL}~b).

The eigenvalue $\m_n(t),n\in\N$, of
the operator $\cL_{dir}(\p(\cdot+t))$ moves
around the interval $(r_n^-,r_n^+)$ making exactly $n$ rounds,
as $t$ runs the interval $[0,1]$,
on the corresponding Riemann surface $\cR$, see \cite{McK81}.
Respectively, the eigenvalue $\gm_n(t)$ of the Dirichlet problem for
the Schr\"odinger operator with the potential $V(E,\p(\cdot+t))$ moves
around the interval $(E_n^-,E_n^+)$ making exactly $n$ rounds,
as $t$ runs the interval $[0,1]$,
on the corresponding Riemann surface $\cE$, see Fig.~\ref{FigrsZEL}.

\begin{figure}
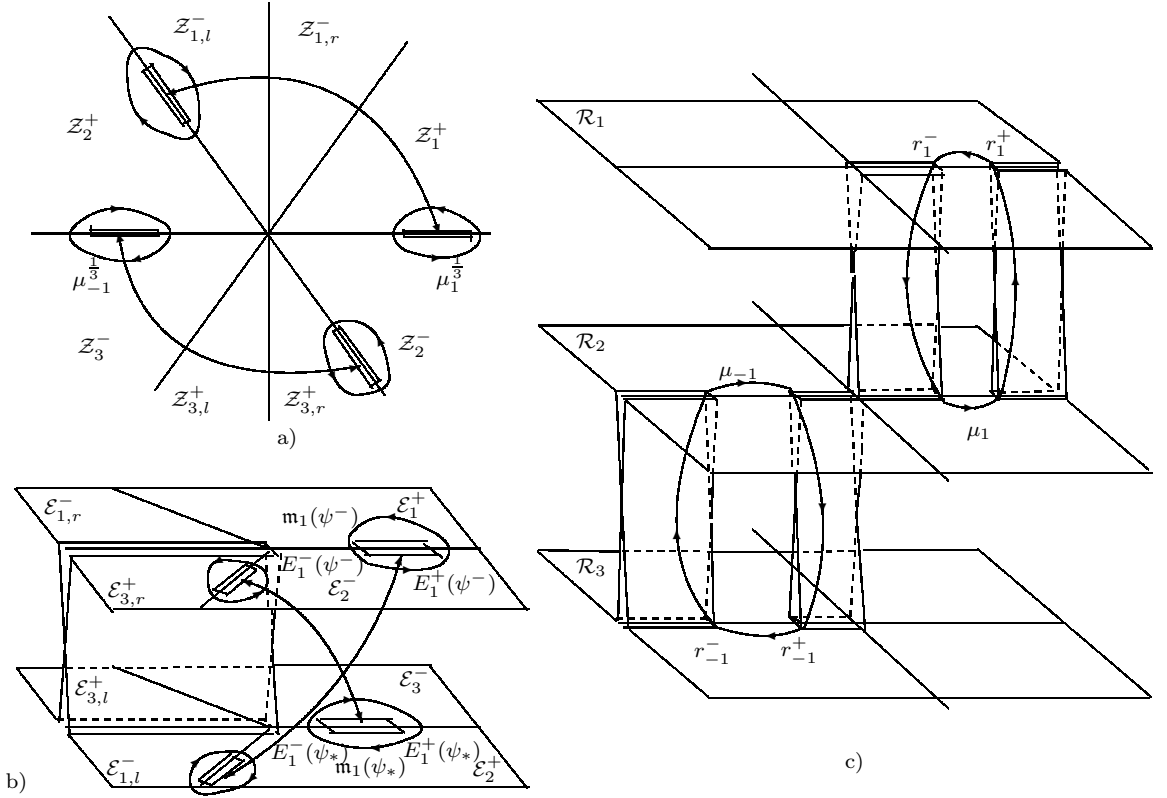

\tiny
\unitlength 0.8mm
\linethickness{0.4pt}
\ifx\plotpoint\undefined\newsavebox{\plotpoint}\fi 

\caption{\footnotesize a) The multipliers $z$-plane, b) $E$-surface $\cE$,
and c) $\l$-surface $\cR$ for the
case $\p$ small. For simplicity we assume here $r_0^+=r_0^-=0$.
On the $z$ and $E$-plane (fig.~a,b) we identify with each other
the corresponding edges of the slits connected in the figure
by curves with arrows: we identify the upper edge of the first slit
with the upper edge of the second slit, we identify the lower edge
of the first slit with the lower edge of the second slit.
The eigenvalue $\m_n(t)$ ($\m_{-n}(t)$) of the
the operator $\cL_{dir} (\p(\cdot+t))$ moves
around the interval $(r_n^-,r_n^+)$ ($(r_{-n}^-,r_{-n}^+)$) making exactly $n$ round,
as $t$ runs the interval $[0,1]$.
}
\lb{FigrsZEL}
\end{figure}

\bigskip

\no\small {\bf Acknowledgments.}
The work was supported by the RSF grant number 23-21-00023.

\end{document}